\documentclass[authoryear,preprint,review,12pt]{elsarticle}

\usepackage{graphicx}
\usepackage[usenames]{color}
\usepackage{natbib}
\usepackage{appendix}

\usepackage{amssymb}
\usepackage{amsmath}
\journal{Icarus}

\usepackage{amsmath,amssymb}

\newcommand\newtext[1]{{\color{black} #1}}

\newcommand\unit[1]{\ \mathrm{#1}}

\def\mm{\unit{mm}}
\def\cm{\unit{cm}}
\def\m{\unit{m}}
\def\km{\unit{km}}
\def\AU{\unit{AU}}

\def\gm{\unit{g}}
\def\My{\unit{My}}
\def\Gy{\unit{Gy}}

\def\s{\unit{s}}

\def\px{\unit{px}}
\def\MPa{\unit{MPa}}

\makeatletter
\def\ps@pprintTitle{%
 \let\@oddhead\@empty
 \let\@evenhead\@empty
 \def\@oddfoot{\footnotesize\itshape
       Submitted to Icarus: January 10,2014\\
		 Revised: August 21, 2014 }%
 \let\@evenfoot\@oddfoot}
\makeatother

\begin{document}

\begin{frontmatter}



\title{Re-examining the main asteroid belt as the primary source of ancient lunar craters.}

\author{David A. Minton\fnref{label1}}
\author{James E. Richardson\fnref{label2}}
\author{Caleb I. Fassett\fnref{label3}}
\address[label1]{Purdue University Department of Earth, Atmospheric, \& Planetary Sciences, 550 Stadium Mall Drive, West Lafayette, IN 47907}
\address[label2]{Arecibo Observatory, Arecibo, PR 00612}
\address[label3]{Department of Astronomy, Mount Holyoke College, South Hadley, MA 01075}

\begin{abstract}
It has been hypothesized that the impactors that created the majority of the observable craters on the ancient lunar highlands were derived from the main asteroid belt in such a way that preserved their size-frequency distribution [Strom, R.G. Malhotra, R., Ito, T. Yoshida, F., Kring, D.A. (2005) Science 309 1847-1850].
\newtext{A more limited version of this hypothesis}, dubbed the E-belt hypothesis, postulates that a destabilized contiguous inner extension of the main asteroid belt produced a bombardment limited to those craters younger than Nectaris basin [Bottke, W.F., Vokrouhlick{\'y}, D., Minton, D., Nesvorn{\'y}, D., Morbidelli, A., Brasser, R., Simonson, B., Levison, H.F. (2012) Nature 485 78-81]. 
We investigate these hypotheses with a Monte Carlo code called the Cratered Terrain Evolution Model ({\em CTEM}), which models the topography of a terrain that has experienced bombardment due to an input impactor population.
We detail our effort to calibrate the code with a human crater counter. 
We also take advantage of recent advances in understanding the scaling relationships between impactor size ($D_{i}$) and final crater size (\newtext{$D_c$}) for basin-sized impact craters (\newtext{$D_c>300\km$}) in order to use large impact basins as a constraint on the ancient impactor population of the Moon.
We find that matching the observed number of lunar highlands craters with \newtext{$D_c\simeq100\km$} requires that the total number of impacting asteroids \newtext{with $D_i>10\km$ be no fewer than $4\times10^{-6}\km^{-2}$}.
However, this required mass of impactors has \newtext{$<1\%$} chance of producing only a single basin larger than the $\sim1200\km$ Imbrium basin; instead, these simulations are likely to produce more large basins than are observed on the Moon.
This difficulty in reproducing the lunar highlands cratering record with a main asteroid belt SFD arises because the main belt is relatively abundant in the objects that produce these ``megabasins'' that are larger than Imbrium. 
We also find that the main asteroid belt SFD has \newtext{$<16\%$} chance of producing Nectarian densities \newtext{of $D_c>64\km$ craters} while not producing a crater larger than Imbrium, as required by the E-belt hypothesis.
These results suggest that the lunar highlands were unlikely to have been bombarded by a population whose size-frequency distribution resembles that of the currently observed main asteroid belt.
We suggest that the population of impactors that cratered the lunar highlands had a somewhat similar size-frequency distribution as the modern main asteroid belt, reflecting a similar rocky composition and collisional history, but had a smaller ratio of objects capable of producing megabasins compared to objects capable of producing $\sim100\km$ craters.
\end{abstract}


\end{frontmatter}

\section{Introduction}
\label{sec:intro}
\defcitealias{CraterAnalysisTechniquesWorkingGroup:1979cs}{Crater Analysis Techniques Working Group, 1979}

The heavily-cratered highlands of the Moon are uniquely suited as a terrain for studying the early bombardment history of the inner solar system.
Little post-bombardment geological processing has modified highlands craters over the $\sim 3.5\Gy$ since the majority of them formed.
This is in contrast to the heavily cratered terrains of Mars and Mercury which are much more heavily modified by geologic processes other than cratering.
We also have at present a substantial sample collection from the Moon, including highlands rocks, thanks to both the lunar meteorite collection \newtext{and} the Apollo and Lunar programs~\citep{Vaniman:1991vi}.
The combination of impact-dominated geology and an abundant and diverse sample record means that the lunar cratering record has been the primary source for absolute-age crater chronology systems.
It is also the best record currently available for testing models for the bombardment history of the inner solar system.

It has been hypothesized that the ancient lunar highlands craters were formed by impacts from a population that was identical to that of the main asteroid belt, based on similarities between the size-frequency distribution of asteroids and the size-frequency distribution of lunar highlands impactors derived using a crater size scaling relationship~\citep{Strom:2005fo}. 
This correlation between the lunar highlands impactors and the main asteroid belt was interpreted to suggest that some \newtext{mass}-independent process perturbed main belt asteroids onto terrestrial planet-crossing orbits.
Because gravitational accelerations are \newtext{mass}-independent~\citep{Galilei:1638tw}, it was hypothesized that the sweeping of mean motion and secular resonances by the migration of giant planets~\citep{Fernandez:1984iy,Malhotra:1993kd,Hahn:1999eh,Tsiganis:2005dg} was responsible for generating the impactor populations that cratered the lunar highlands.
Dating of lunar samples and evidence from geological superposition suggests at least two large lunar basins (where ``basin'' here refers to a crater with final rim-to-rim diameter \newtext{$D_c>300\km$}), Imbrium and Orientale, were formed $600$--$800\My$ after solar system formation~\citep{Wilhelms:1987vb}, and therefore the migration of the giant planets needed to be delayed by that amount of time in order to produce at least these basins from the asteroid belt.

A scenario for giant planet orbital evolution dubbed the ``Nice model'' was developed, in which the giant planets initially formed in a tightly packed, quasi-stable configuration with a massive icy planetesimal disk beyond Neptune that was disrupted when Jupiter and Saturn crossed a mean motion resonance~\citep{Tsiganis:2005dg,Morbidelli:2005dr,Gomes:2005gi}.
The Nice model provides a mechanism to allow a delay between planet formation and the onset of the Late Heavy Bombardment (LHB).
It was estimated that the Nice model could have produced a bombardment of $\sim10^{22}\gm$ of asteroids onto the Moon, plus an equivalent amount of comets~\citep{Gomes:2005gi}.
This mass is enough to produce all of the of observed lunar basins on the Moon~\citep[c.f.][]{Ryder:2002gd}.
\newtext{Assuming an asteroid mean density of $\rho_i=2.5\gm\cm^{-3}$ and using a size-frequency distribution of asteroids obtained by the Wide-field Infrared Survey Explorer spacecraft (WISE)~\citep{Masiero:2011jc}, this corresponds to a number density of $D_i>10\km$ impactors of approximately $9.7\times10^{-7}\km^{-2}$.}

However, the original Nice model calculations for the magnitude of the lunar bombardment produced by disrupting the main asteroid belt was challenged when it was shown that if the migration rates of Jupiter and Saturn were too slow they would have caused the terrestrial planets to become disrupted~\citep{Brasser:2009ce,Agnor:2012bo}.
Also, a migration rate that was too slow would have altered the distribution of asteroid eccentricities~\citep{Minton:2011ir} and inclinations~\citep{Morbidelli:2010kj} in ways that are not supported by observation.
A more rapid evolution of Jupiter and Saturn compatible with these inner solar system constraints occurs in a small subset of Nice model simulations dubbed the ``Jumping Jupiter'' scenario when the ice giants (Uranus and Neptune) have close encounters with Jupiter and Saturn~\citep{Brasser:2009ce}.
However, a more rapid migration of the giant planets would not have produced the required amount of mass needed to crater the lunar highlands and produce all of the observed lunar basins.
It was hypothesized that the main asteroid belt may have had a primordial inner extension, dubbed the E-belt, which is a region that would only have become unstable after giant planet migration occurred and the $\nu_6$ secular resonance arrived at $\sim2\AU$~\citep{Bottke:2012ce}.
The destabilized E-belt, along with a small contribution from the main asteroid belt, would produce $\sim1/3$ of the lunar basins as a result of rapid giant planet migration under the Jumping Jupiter scenario, or the $\sim14$ youngest basins beginning with Nectaris~\citep{Fassett:2012hr}.
The more limited view of the LHB suggested by the E-belt model postulates that the pre-Nectarian terrains may have been created by leftover planetesimals from the epoch of terrestrial planet formation~\citep{Morbidelli:2012ko}, and possibly also include a contribution from cometary impactors, which should dominate total impactor mass~\citep{Gomes:2005gi}, although it is not clear why a non-asteroidal signature is not apparent in the lunar cratering record~\citep{Bottke:2012ce}.
\newtext{Putative impactor remnants identified in lunar samples also favor asteroids over comets in both highly siderophile elements~\citep{Kring:2002ip} and oxygen isotopes \citep{Joy:2012kt}.}
A more extensive review of this topic may be found in~\cite{Fassett:2013cd}.

Here we revisit the hypothesis that the lunar highlands were cratered by main belt asteroids.
A powerful Monte Carlo code for studying the evolution of cratered terrains has recently been developed~\citep{Richardson:2009dc}.
The code is named, simply enough, the Cratered Terrain Evolution Model ({\em CTEM}). 
We used {\em CTEM} to model the cratering history of the lunar highlands by bombarding simulated lunar surfaces with a main belt asteroid impactor population.
Our goal is to quantify the amount of cratering required to reproduce the observed lunar highlands crater record.
We do this by performing a large number of {\em CTEM} simulations of impactors bombarding a lunar-like surface and then comparing the outcomes with published crater counts for the lunar highlands.
For this study we use a catalog of craters produced by the Lunar Orbiter Laser Altimeter (LOLA) aboard the Lunar Reconnaissance Orbiter spacecraft~\citep{Head:2010fy}.
Our impactor population uses the most up-to-date size-frequency distribution (SFD) for the main asteroid belt as obtained by the Wide-field Infrared Survey Explorer (WISE)~\citep{Masiero:2011jc}.
\newtext{For impactor sizes below $5\km$, where WISE data becomes unreliable due to biases, we use the main belt SFD from the Sub-Kilometer Asteroid Diameter Survey (SKADS)~\citep{Gladman:2009cx}, assuming $9\%$ albedo.}

This paper is divided into two main sections.
In the first section we will describe in detail our methods, including a description of the {\em CTEM} code as well as our efforts to calibrate it with a human crater counter.
In the second section, we will describe the results of applying the {\em CTEM} model to the problem of the impact history of the lunar highlands. 

\section{Methods}
\label{sec:methods}
For this project we employ a \newtext{Monte Carlo} cratering code called the Cratered Terrain Evolution Model ({\em CTEM})~\citep{Richardson:2009dc}.  
Numerous improvements have been made to {\em CTEM} since the \cite{Richardson:2009dc} study, which we will describe in detail here.
First, several performance improvements have been made to the code, including the addition of parallelization, allowing us to explore a much wider parameter space for a given study than was available previously.
The code was also re-written in Fortran 2003, and allows a much greater degree of extensibility and modularity than the earlier Fortran 77 version.
Lastly, we have significantly revised the crater-counting algorithm within the code and performed a comprehensive calibration study with a human crater counter (co-author Fassett), allowing us to better apply the computer model results obtained from {\em CTEM} to the interpretation of remotely observed planetary surfaces.

{\em CTEM} models a given planetary surface as a grid of points (heretofore referred to as pixels), where the number of pixels in the grid is set by the user.
Each pixel contains a variety of data, one of which is the current elevation at that location, allowing the construction of a Digital Elevation Model (DEM) of the surface.
For each impact that is modeled, the DEM of the surface is altered to represent the topographic changes brought about in the formation of a crater.
Because {\em CTEM} models the topography of craters, rather than idealizing them as simple shapes such as rings or circles, a number of important crater degradation processes are naturally produced by the code, including unstable slope collapse, impact ejecta, and diffusive erosion.

Determining whether a crater is countable or not is done by measuring how much the topography of a given crater deviates from its original shape.
In order to quantify crater ``countability'' within the code, we performed a crater counting calibration study with {\em CTEM} generated DEMs by supplying them to a human crater counter (co-author Fassett), who analyzed {\em CTEM}-generated DEMs in a \newtext{way} similar to how craters were counted using LOLA-generated lunar DEMs.
Therefore the crater counting algorithm of the {\em CTEM} code has been calibrated using the same process that our comparison data set was generated from.

Here we summarize all of the physical processes that {\em CTEM} models, and further details regarding the code's capabilities are described in \cite{Richardson:2009dc}.
Once an impactor size, velocity, and impact angle are determined, the following sequence of steps is performed to model the resulting crater and its effects on the topography of the simulated terrain:
\begin{enumerate}
\item Perform a crater scaling calculation to determine the dimensions of the resulting crater.
\item Place the crater at a random location on a surface and modify the terrain to reflect the final crater shape. We use a repeating boundary condition to mitigate against edge effects.
\item Emplace ejecta onto the surface and degrade the pre-existing terrain affected by ejecta deposition.
\item Collapse any slopes that are above the angle of repose.
\end{enumerate}
Once at the end of a run, as well as periodically throughout the run, {\em CTEM} will examine the simulated surface and tally up the number of observable craters.
In Subsection~\ref{subsec:scaling} we will we will describe the crater scaling relationship used by {\em CTEM} as well as its method for emplacing ejecta onto the surface.
Because including estimates of the number of large lunar impact basins is a critical component of our present work, in Subsection~\ref{subsec:basins} we describe how we incorporated recently published scaling relationships for lunar basins based on computer hydrocode modeling into {\em CTEM}.
Next, in Subsection~\ref{subsec:degradation} we will describe a number of ways that craters may degraded over time on a {\em CTEM}-generated surface.
Then, in Subsection~\ref{subsec:calibration} we will describe the study that we performed in order to calibrate the crater recognition algorithm within {\em CTEM} with a human crater counter.
Finally, in Subsection~\ref{subsec:impactor_population} we will describe the impactor population model that we adopted for this study.

\subsection{Crater and ejecta scaling model}
\label{subsec:scaling}
The impact crater volume scaling relationships that we use to relate the size of a primary impactor to the size of its resulting crater on a particular target surface, given a set of projectile and target material properties and impact parameters, are reviewed in \cite{Holsapple:1993js}.  
Previously, most applications of such relationships dealt strictly in either the gravity- or strength-dominated cratering regimes; for example, \cite{Vickery:1986hz}. 
However, cratering on small scales or on small target bodies often falls into neither regime:  gravity and target strength are both important to the size of the final crater~\citep{Ivanov:2001vc,Richardson:2005io}.  
We have therefore adopted the general solution to the transient crater volume scaling relationship given by Eq.~19 in \cite{Holsapple:1993js}, which includes both gravity and strength terms. 

The application of a general solution to the crater volume scaling relationship permits us to also utilize the general solution to the ejecta velocity scaling relationships developed in \cite{Richardson:2007ce}. 
These revised ejecta scaling relationships permit us to directly compute surface ejection velocity as a function of projectile and target material properties, impact parameters, and distance from the impact site, all the way out to the transient crater rim.
This analytical solution is applied to a discretized model (by a factor of 10$^3$ to 10$^4$) of the excavation flow-field's hydrodynamic streamlines~\citep{Maxwell:1974tk, Maxwell:1977ud}, in order to compute excavated mass as a function of distance from the impact site and ejection velocity.
Finally, these discrete ejected mass segments are followed in post-ejection flight through a set of ballistics equations in order to compute ejecta blanket thickness as a function of distance from the final crater rim, as described in \cite{Richardson:2009dc}.  
This scaling-relationship based excavation-flow properties model is quite general in nature, and can be applied to craters ranging in size from the laboratory \citep{Richardson:2011bq} to impacts on small solar system bodies \citep{Richardson:2007ce} to large, planetary scales \citep{Richardson:2009dc}.
This allows us to model ejecta deposition over a wide range of gravity and projectile/target material properties and impact parameters.  

The principal equations utilized in this impact ejecta model are as follows (from the Appendix to \cite{Richardson:2011bq}).
The central analytical expression of the ``excavation flow properties model'' describes the effective particle ejection velocity $v_{ef}$ as a function of radial distance $r$ from the impact site as follows (from Secs.~2.2 \& 2.3 of \cite{Richardson:2007ce}):
\begin{equation}
\label{bernoulli_final}
v_{ef}(r) = \left[ v_e^2 - C_{vpg}^2 g r - C_{vps}^2 \frac{\bar{Y}}{\rho_t} \right]^{\frac{1}{2}} \>,
\end{equation}
where $\rho_t$ is the target density, $g$ is the gravitational acceleration, and $\bar{Y}$ is the effective target strength with respect to crater excavation. The uncorrected ejection speed $v_e$ is given by the gravity-scaled ejecta speed equation derived by \cite{Housen:1983kp}:
\begin{equation}
\label{velposition_g}
v_e(r) = C_{vpg} \sqrt{g R_g} \left( \frac{r}{R_g} \right)^{- \frac{1}{\mu}} \>.
\end{equation}
where $R_g$ is the gravity-scaled transient crater radius; that is, the radius that would result if the target was strengthless.  $\mu$ is a commonly used exponential material constant~\citep{Holsapple:1993js}, and the constant $C_{vpg}$ is given in \cite{Richardson:2007ce} as:
\begin{equation}
\label{constant_velposition_g}
C_{vpg} =  \frac{\sqrt{2}}{C_{Tg}} \left( \frac{\mu}{\mu + 1} \right) \>.
\end{equation}
This expression contains constant $C_{Tg}$, which has an experimentally determined value of 0.8--0.9~\citep{Schmidt:1987vc, Holsapple:2007bd} as discussed in Sec.~2.2 of \cite{Richardson:2007ce}.

The constant $C_{vps}$ in the strength term of Eq.~\ref{bernoulli_final} is given by:
\begin{equation}
\label{constant_velposition_s_2}
C_{vps} = C_{vpg} \left[ \frac{\rho_t g R_g}{\bar{Y} + Y_t} \right]^{\frac{1}{2}} \left( \frac{R_g}{R_s} \right)^{\frac{1}{\mu}} \>,
\end{equation}
where $R_s$ is the transient crater radius due to the effects of both gravity and strength.  The transition strength $Y_t$ (between gravity and strength dominated cratering) is expressed as:
\begin{equation}
\label{transition_strength}
Y_t = \rho_t v_i^2 \left[ \left( \frac{g a_i}{v_i^2} \right) \left( \frac{\rho_i}{\rho_t} \right)^{\frac{1}{3}} \right]^{\frac{2}{2 + \mu}} \>.
\end{equation}
where $v_i$ is the vertical component of the impactors speed, $a_i$ is the impactor's mean radius, and $\rho_i$ is the impactor density.

In order to estimate the size of the transient crater produced by a particular impact, we make use of the general solution to the crater-size scaling relationship developed in~\cite{Holsapple:1993js}, which contains the effects of both gravity and strength:
\begin{equation}
\label{volume_final}
V = K_1 \left( \frac{m_i}{\rho_t} \right) \left[ \left( \frac{g a_i}{v_i^2} \right) \left( \frac{\rho_t}{\rho_i} \right)^{- \frac{1}{3}} + \left( \frac{\bar{Y}}{\rho_t v_i^2} \right)^{\frac{2 + \mu}{2}} \right]^{- \frac{3 \mu}{2 + \mu}} \>,
\end{equation}
where $K_1$, $\mu$, and $\bar{Y}$ are experimentally derived properties of the target material.  
We adopt $K_1=0.22$ and $\mu=0.55$ for our study, as these are expected values for hard rock target material based on experiments~\citep{Holsapple:1993js}.
In practice, the effective target strength $\bar{Y}$ is roughly a measure of ``cohesion," and usually lies somewhere between the laboratory-measured tensile and shear strengths of the material \citep{Holsapple:1993js, Holsapple:2007bd}, for low- to medium-porosity targets.  
In high-porosity targets, the effective target strength $\bar{Y}$ can be thought of in broader terms:  as the energy per unit volume (rather than the force per unit area) required to both crush pore spaces and break the material apart for excavation.
Because in our study we are only comparing our model against \newtext{$D_c>20\km$} craters, our scaling relationship is insensitive to effective target strength, $\bar{Y}$.

The transient crater volume $V$ can be related to the more easily measured transient crater (rim-to-rim) diameter $D_t$ or radius $R_t$ by:
\begin{equation}
\label{volume_diam_rad}
V = \frac{1}{24} \pi D_t^3 = \frac{1}{3} \pi R_t^3 \>,
\end{equation}
where we assume that the transient crater depth $H_t$ is roughly $1/3$ its diameter $D_t$: in experiments this is somewhat variable, between $1/4$ and $1/3$~\citep{Schmidt:1987vc, Melosh:1989uq}.

Note that the above expressions yield only a transient crater size; that is, the crater's momentary diameter prior to gravitational collapse.  In order to compute a final crater diameter \newtext{$D_c$} from the transient crater diameter $D_t$, two expressions are used, one for small, simple craters, and one for large, complex craters --- see the discussion in \cite{Chapman:1986vj} for more details.  For simple craters, the transient crater is approximately $80$--$85\%$ the diameter of the final crater, and the conversion factor is linear:
\begin{equation}
\label{simple_crater_final}
D_s = 1.25 D_t \>,
\end{equation}
as described in \cite{Chapman:1986vj} and \cite{Melosh:1989uq}.  For complex craters the conversion follows a power-law, as described in \cite{Croft:1981uq}, \cite{Chapman:1986vj}, \cite{Melosh:1989uq}, and \cite{Schenk:2004bt}.  This power-law begins at a simple crater diameter known as the simple-to-complex transition point $D_{tr}$, and where this transition point varies inversely with the surface gravity of the body involved ($D_{tr} \propto 1/g$).  

Fitting the data shown in Fig.~12.18 of \cite{Schenk:2004bt}, we developed and expression for the simple-to-complex transition for silicate rock:
\begin{equation}
\label{silicate_transition}
D_{tr} = 16533.8 g^{-1.0303} \>,
\end{equation}
where the gravity $g$ is given in terms of m~s$^{-2}$, and the transition crater diameter $D_{tr}$ is given in term of meters.

In addition to the above method for determining the diameter at which simple craters transition to complex craters, we also need a method for computing the size of the final complex crater given the transient crater diameter from Eq.~\ref{volume_diam_rad}.  
For a silicate rock body, the final diameter of a complex crater is given by:
\newtext{
\begin{equation}
\label{silicate_final}
D_c = D_s \cdot \frac{D_s}{D_{tr}}^{0.179} \>,
\end{equation}
}
where the simple crater diameter $D_s$ is computing using Eq.~\ref{simple_crater_final}, and the above expression is used only when $D_s > D_{tr}$; for simple craters \newtext{$D_c=D_s$.}  
\newtext{The silicate body power-law exponent is taken from \cite{Croft:1985uu}.}

\newtext{Finally, we set a maximum depth for the floors of complex craters to a depth, $h_{max}$ given by \cite{Pike:1977up}:
\begin{equation}
h_{max}=1.044D_c^{0.301}.
\label{e:Pikedepth}
\end{equation}}
This creates complex craters with a characteristic flat-floored shape.
The current version of {\em CTEM} does not attempt to model central peaks or peak rings.

\subsection{Scaling relationships for large lunar impact basins}
\label{subsec:basins}
Large lunar basins present a challenge for crater scaling relationships, as the formation of a lunar basin, which we define as any crater with a final diameter \newtext{$D_c > 300\km$}, involves energies far beyond those approachable in laboratory or field studies.
Even the largest known impact craters on Earth are somewhat below the size of a ``small'' lunar basin.
Nevertheless, a number of advancements have been made in recent years in understanding the processes involved in lunar basin formation using numerical hydrocodes.

\cite{Potter:2012gm} developed a scaling relationship for lunar basins based on hydrocode simulations that relates the size of the transient crater to a crustal feature called the ``crustal annulus,'' which is observed in the crust beneath lunar basins~\citep{Wieczorek:1999fh,Hikida:2007fe,Melosh:2013cz}.
The size of a basin's crustal annulus is often easier to determine than its rim-to-rim size.
\cite{Potter:2012gm} also showed showed that the transient crater diameter was sensitive to the thermal profile of the lunar upper mantle, and developed two scaling \newtext{relationships} to account for changes to the lunar thermal gradient over time as well as the major hemispherical dichotomy of the Moon. 
They define two thermal profiles, TP1 and TP2, where TP1 represents a weaker, warmer upper mantle, and TP2 represents a colder, stronger upper mantle. 
Because the near-side crust is thinner and possibly warmer due to higher abundances of radioactive nuclides within the Procellarum KREEP terrain~\citep{Jolliff:2000im}, this thermal dependence means that the size of basins that form on the near side tend to be larger than those on the far side for a given impactor size.
This effect has been seen in data returned by the Gravity Recovery and Interior Laboratory (GRAIL) spacecraft~\citep{Miljkovic:2013bx}
\newtext{\cite{Miljkovic:2013bx} showed that the observation that near side basins are systematically larger than far side basins could be explained by the warmer mantle of the near-side.
When the crustal dichotomy of the thermal gradient was accounted for, the near side and far side basins impactors had an identical SFD.}
The scaling relationships that relate impactor size to crustal annulus size in the simulations of \cite{Miljkovic:2013bx} are consistent with those in \cite{Potter:2012gm}.

In {\em CTEM} we calculate the size of the transient crater for a given impactor using Eqs.~\ref{volume_final} and \ref{volume_diam_rad}.
Using the results of \cite{Potter:2012gm}, we can then relate the diameter of the crustal annulus, $D_{ca}$, to our calculated transient crater diameter, $D_t$.
The warm/weak TP1 scaling relationship is given in terms of the crustal annulus diameter:
\begin{equation}
D_{ca,TP1} = 0.0469D_{t}^{1.613} 
\label{e:DcaTP1}
\end{equation}
where diameters are given in units of km.
The cold/strong TP2 scaling relationship is given as:
\begin{equation}
D_{ca,TP2} = 0.103D_{t}^{1.389} 
\label{e:DcaTP2}
\end{equation}

Although {\em CTEM} has the capability of modeling crustal thickness, our crater counting algorithm is calibrated for surface topography, and therefore we  need an estimate of the final crater rim. 
\cite{Potter:2012gm} do not explicitly relate the size of the crustal annulus to the size of the final rim, however they do provide measurements for three well-studied craters, Orientale, Imbrium, and Serenitatis.
Imbrium was modeled using TP1, and Orientale was modeled using the TP2.
The Orientale final rim diameter of $930\km$ used in our crater catalog is $\sim56\%$ larger than the crustal annulus.
Imbrium and Serenitatis rims are $\sim30\%$ larger and $\sim10\%$ larger than their respective crustal annuli. 
Therefore to obtain the final crater rim diameter, \newtext{$D_c$}, from the crustal annulus diameter, $D_{ca}$:
\newtext{
\begin{equation}
D_c=
\begin{cases}
	1.2D_{ca} &: TP1\\
	1.56D_{ca} &: TP2 
\end{cases}
\label{e:Dca2Df}
\end{equation}
}

Because of the statistical nature of our study, we choose to apply either the TP1 or TP2 scaling relationship for a given crater with a $50\%$ probability to reflect the hemispherical dichotomy of the Moon.
We apply our basin scaling relationship for impactors with $D_i>35\km$.
To test our basin scaling relationship, we generated a test crater the size of Orientale.
Because the ejecta blanket produced by {\em CTEM} is determined by the excavation flow within the transient crater as well as the distance of the final crater rim from the transient crater~\citep{Richardson:2009dc}, the thickness profile of the ejecta blanket of Orientale acts as an additional constraint on our basin scaling parameters.

\cite{Fassett:2011dj} estimated the thickness profile of Orientale proximal ejecta by counting small craters near the rim of Orientale to determine the size dependence of those craters that escaped being buried by ejecta.
They determined that a best fit ejecta thickness profile was given by the power law:
\begin{equation}
h_{e}=2.9\km\left(4/R_{cr}\right)^{-2.8}
\label{e:orientale_ejecta_profile}
\end{equation}
In order to reproduce a crater the size of Orientale with \newtext{$D_c=930\km$} and an ejecta blanket thickness at the rim of $2.9\km$, using cold/strong mantle thermal profile TP2, we require a \newtext{$D_{i}=75\km$} impactor with a velocity of $15\km\s^{-1}$, and scaling parameters of \newtext{$K_1=0.22$} and \newtext{$\mu=0.55$}.
\newtext{This is consistent with the velocity and size of the Orientale impactor of $50$--$80\km$ found by \cite{Potter:2013gw}, using the iSALE hydrocode and constraints from crustal thicknesses derived from GRAIL spacecraft data.
Using a scaling relationship from \cite{Holsapple:1993js}, \cite{Ryder:2002gd} derived a $4\times10^{20}\gm$ impactor traveling at $20\km\s^{-1}$ for Orientale, which corresponds to a $64\km$ impactor for $\rho_i=2.5\gm\cm^{-3}$.
Our adopted scaling relationship for basins therefore errs on the upper end of previously estimated ranges of Orientale-impactors, which makes it a conservative model for our problem.
}
We will discuss the ejecta blanket profile of this test crater in more detail in Section~\ref{subsubsec:ejecta_burial}.

As large basin impacts on the scale of $\sim2500\km$ South Pole-Aitken basin are important for our study, we tested our scaling relationship for these large megabasins.
\cite{Potter:2012dw} found using hydrodcode modeling that the best match to SPA was found for $D_{i}=170\km$ traveling at $10\km\s^{-1}$ and impacting into mantle with a high thermal gradient (more similar to TP1 than TP2).
For this size impactor and velocity, using the TP1 crater scaling relationship we obtain a final diameter of of \newtext{$D_c=2033\km$}, and for TP2 we obtain \newtext{$D_c=1355\km$}.
This impactor size is also consistent with other hydrocode simulations of the formation of SPA~\citep{Collins:2004ua}.
Therefore, {\em CTEM} reproduces basins at the scale of SPA, and in fact we may be somewhat conservative in the size of basins produced by impactors similar to the one that \cite{Potter:2012dw} modeled.

\subsection{Crater degradation modeling with {\em CTEM}}
\label{subsec:degradation}
In order for {\em CTEM} to properly model a surface in or near the state of cratering equilibrium, we must first calibrate the crater detection algorithm to ensure that the craters that are counted by {\em CTEM} are equivalent to the craters that a human would count on a real surface with an equivalent cratering history.
Here we first describe how {\em CTEM} handles different mechanisms of crater erasure, and then describe the work we have done to calibrate {\em CTEM}'s crater identification algorithm with a human crater counter. 

The difficulty in implementing accurate crater obliteration modeling may be restated as the difficulty in identifying ``countable'' model craters in a way that mimic what a human crater counter would also consider countable on real terrain with an equivalent bombardment history.
This difficulty is one that has been encountered by all researchers that have attempted to construct Monte Carlo cratering codes. 
A limitation in past attempts has been the confluence of limited computing power and a problem that requires a great deal of computer resources to study, due to the huge range in spatial scales and the power law nature of impacting populations.
We take advantage of advances in computing power in order to model the cratering process as a sequence of topographic changes produced on a simulated surface represented by a finite grid of elevation points, or digital elevation model (DEM).
This allows us to model the processes of crater formation and obliteration naturalistically.

For the global and regional scale lunar cratering simulations performed for this study, old craters are obliterated by four primary mechanisms, each of which is modeled within the present version of {\em CTEM}: 1) Cookie-cutting, where new craters destroy craters directly by forming on the same terrain, 2) Sandblasting, where numerous small craters erode an old crater enough that the old crater is no longer recognized, 3) Seismic shaking, where the surface vibrations produced by each new impact destabilize and degrade slopes and craters, and 4) Ejecta burial/ballistic sedimentation, where ejecta deposits fill in and erode old craters.
In the following subsections we will describe how {\em CTEM} handles the three most important crater obliteration mechanisms for our study (cookie-cutting, sandblasting, and ejecta burial/ballistic sedimentation). 
We will also describe how we have tested scenarios wherein one of the three mechanisms dominates over the others.
On large bodies, such as the Earth's moon, impact-induced seismic shaking likely plays only a minor role in crater obliteration, particularly on the regional and global scales considered in this particular study~\citep{Houston:1973um}.
\newtext{Therefore we do not employ the seismic shaking model here.}

\subsubsection{Crater erasure mechanisms: Cookie Cutting}
\label{subsubsec:cookie_cutting}
Cookie cutting describes the most direct way that new craters destroy old craters, which is by simple geometric overlap~\citep{Woronow:1977ta,Woronow:1977kf,Woronow:1978kg}.
When an impactor strikes a cratered surface, any pre-existing craters within the final rim of the new crater are obliterated during the crater formation process.
{\em CTEM} implements cookie cutting in a simple way, but in order to describe it we must first digress and explain a feature of the code that facilitates crater identification during the counting stage.

The surface that {\em CTEM} models is represented by a grid of pixels with a repeating boundary.
Every pixel on a {\em CTEM} surface grid contains a variety of information in addition to elevation. 
When {\em CTEM} generates a crater centered at a particular position on the grid, the pixels interior to a circle of diameter \newtext{$1.05D_c$}, where \newtext{$D_c$} is the crater's final rim diameter, are tagged with the crater's unique identifier.
The value of \newtext{$D_c$} is used as the unique identifier for each crater, as it is unlikely that any two craters will have precisely the same diameter (to double precision).
In order for {\em CTEM} to successfully identify topographic depressions as possible craters, the code must allow for any given pixel to contain the tags of multiple craters, otherwise a large crater could become completely erased by multiple small impacts despite the original crater being identifiable as as topographic depression.
We implement this capability in a layering system, where a single pixel contains multiple crater tags.
The existence of a crater's tag on the grid is a necessary, but not sufficient, condition for it to be counted.

\begin{figure}[htb] 
\centering
\includegraphics[width=\textwidth]{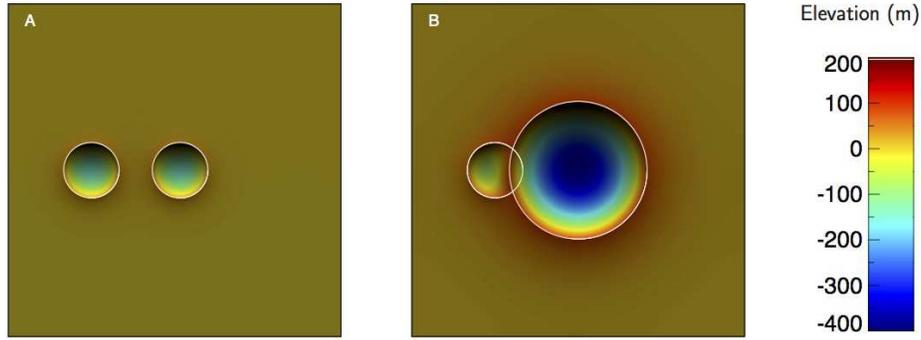} 
\caption{Example of crater degradation by ``cookie cutting.''
A) Two test craters are placed onto a fresh surface.  
The impactor parameters were: $D_i=76.7\m$, $v_i=15\km\s^{-1}$, $\rho_i=2.5\gm\cm^{-3}$. 
The final crater diameters were \newtext{$D_c=1.0\km$}.
One crater was placed at the center of the grid, and the second one was placed $100\m$ to the left of center. 
Two craters are counted by {\em CTEM}, indicated by being circled in white.
B) The surface shown in A) after a larger crater has formed in the center of the grid. 
The impactor parameters were: $D_i=200\m$, $v_i=15\km\s^{-1}$, $\rho_i=2.5\gm\cm^{-3}$. 
The final crater diameter was \newtext{$D_c=2.48\km$}. 
The new crater has completely obliterated the older, smaller centered crater, but only partially cut the smaller offset crater.
{\em CTEM} no longer counts the $1\km$ centered crater, but still counts the partially cut offset crater.
The grid was $1000^2\px$, and the resolution was $6\m\px^{-1}$. 
Lunar gravity was assumed as $g=1.62\m\s^{-2}$ and target density was $\rho_t=2.7\gm\cm^{-3}$.}
\label{f:cookiecut}
\end{figure}

If pre-existing craters within the new crater's final rim are deemed to be removed due to cookie cutting, then the tag that identifies the old crater is removed from the pixel.
The cookie cutting mechanism generally only applies if the new crater is comparable in size or larger than the pre-existing craters. 
So when a pixel is tagged with a new crater, the tags for all craters smaller than the new crater within the region where the two craters overlap are removed.
We show an example of craters affected by cookie cutting in Fig.~\ref{f:cookiecut}.
In that figure we show two small test craters that are affected by the formation of a larger crater.
The larger crater completely overlaps one of the smaller craters, and removes all of its tags rendering it uncountable. 
However, it only partially overlaps the other of the smaller craters, and only removes some of its tags. 
{\em CTEM} then evaluates the topography of the remaining pixels and determines that the crater is still countable (see Section~\ref{subsec:calibration}).

Small craters that form inside of larger ones don't necessarily erase the larger ones.
A large crater can become completely covered in small craters, and still be recognized as a topographic depression and counted, since the smaller craters did not remove the larger crater's tags.
Pre-existing craters that are larger than newer overlapping craters can only be removed by substantial topographic alteration.

\subsubsection{Crater erasure mechanisms: Sandblasting}
\label{subsubsec:sandblasting}
The term ``sand-blasting'' has been used to describe the process by which many small impacts can erode the surfaces of asteroids~\citep{Chapman:1976hj}, but was initially adopted for the analogous process by which numerous small craters erode larger craters to the point where the larger craters are not recognizable as a crater by a human crater counter~\citep{Soderblom:1970cn,Hartmann:1984hb}.
Human crater counters regularly identify topographic depressions on planetary surfaces as degraded craters. 
Therefore, in order for {\em CTEM} to count as craters those features which would be counted by a human, it must be able to successfully identify sandblasted craters.

\begin{figure}[b] 
\centering
\includegraphics[width=\textwidth]{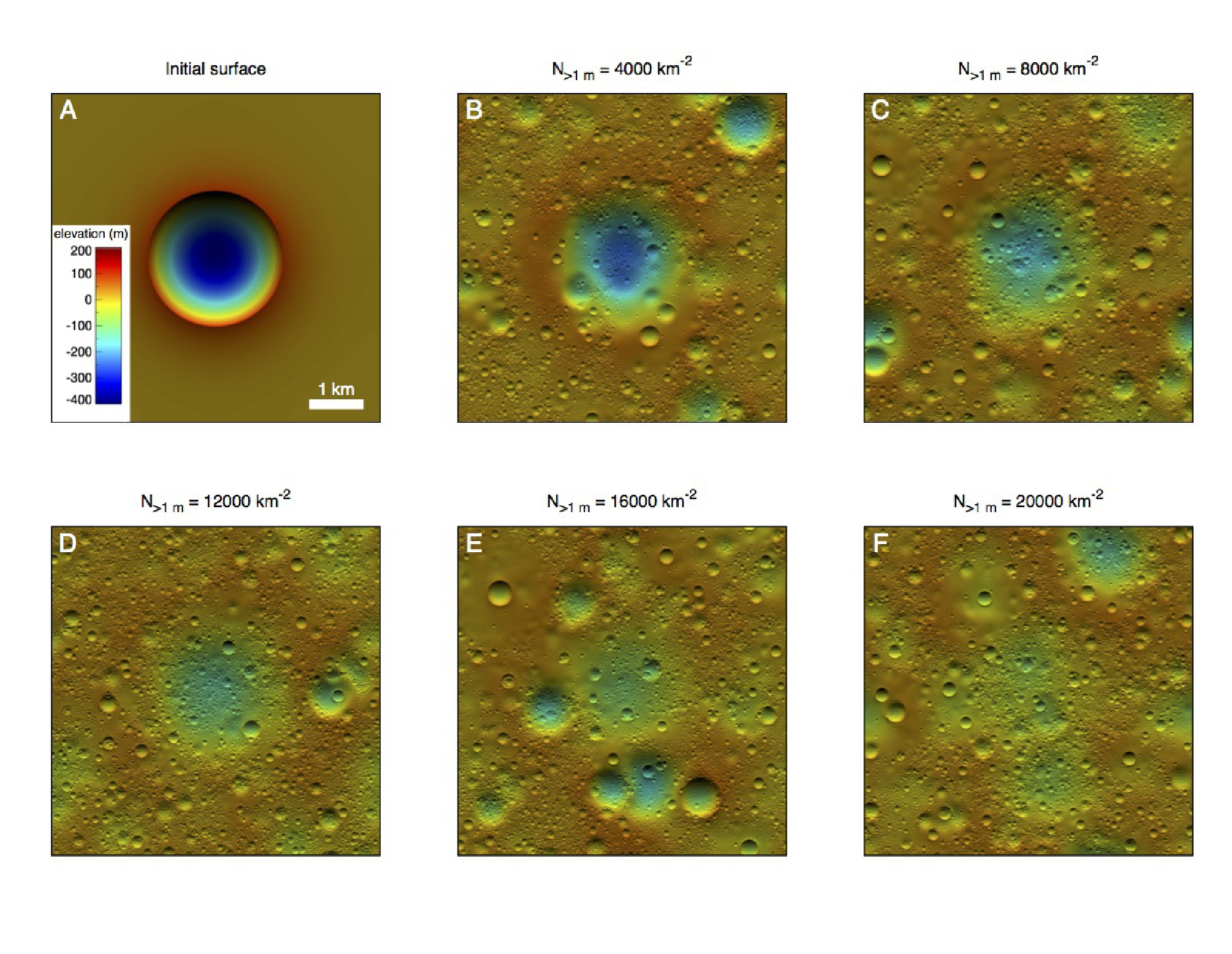} 
\caption{This series of images shows a sequence of outputs from a {\em CTEM} run demonstrating crater obliteration via sandblasting.
A) The test crater shown with \newtext{$D_c=2.5\km$} is placed on a fresh lunar surface.
B--F) Each frame in this sequence shows the topography of the simulated surface after the accumulation of \newtext{$N_{>1\m}=4000\km^{-1}$}. 
The impactors have a size-frequency distribution of $N_{>D}\propto D^{-3}$ and a velocity distribution similar to asteroidal impactors on the Moon with an RMS value of $18.3\km\s^{-1}$~\citep{Yue:2013bx}.
The grid was $1000^2\px$, and the resolution was $6\m\px^{-1}$. 
Lunar gravity was assumed as $g=1.62\m\s^{-2}$, target density was $\rho_{t}=2.7\gm\cm^{-3}$, and projectile density was $\rho_p=2.5\gm\cm^{-3}$.
}
\label{f:sandblast-sequence}
\end{figure}

When multiple small craters impact the surface occupied by a larger old crater, the new craters transport ejected material and, over time, the topography of the large old crater relaxes back toward the mean elevation of the terrain.
This topographic diffusion is seen to occur on the lunar mare~\citep{Fassett:2013ue}.
This process can be seen clearly in a {\em CTEM} simulation as shown in the sequence of images in Fig.~\ref{f:sandblast-sequence}.
We begin with a $2.5\km$ test crater and bombard the surface with an impactor population with a size-frequency distribution described as $N_{>D}\propto D^{-3}$.
The impactors have a velocity distribution similar to asteroidal impactors on the Moon with an RMS of $18.3\km\s^{-1}$~\citep{Yue:2013bx}.
As the total number of impacts accumulates, the surface becomes visibly degraded and at some point the original crater shown in Fig~\ref{f:sandblast-sequence}a might no longer be countable by a human observer.

\subsubsection{Crater erasure mechanisms: Ejecta burial}
\label{subsubsec:ejecta_burial}

The ejecta from a fresh crater can obliterate old craters beyond its rim by burying them in ejecta.
Intuitively, it seems obvious that if the ejecta thickness is greater than the depth of a crater, that the crater will be buried, and thus will be invisible to a crater counter.
This concept has been employed to estimate the thickness of ejecta deposits from lunar craters~\citep{Moore:1974vy}
Recently, \cite{Fassett:2011dj} measured the thickness of the ejecta blanket of the \newtext{$D_c=930\km$} Orientale basin by measuring the rim-to-bowl depth of craters as a function of distance from the basin's rim (taken to be the Cordillera Mountain ring) and comparing the results to the expected depth of the craters, based on crater scaling relationships.

\begin{figure}[htb] 
\centering
\includegraphics[width=0.9\textwidth]{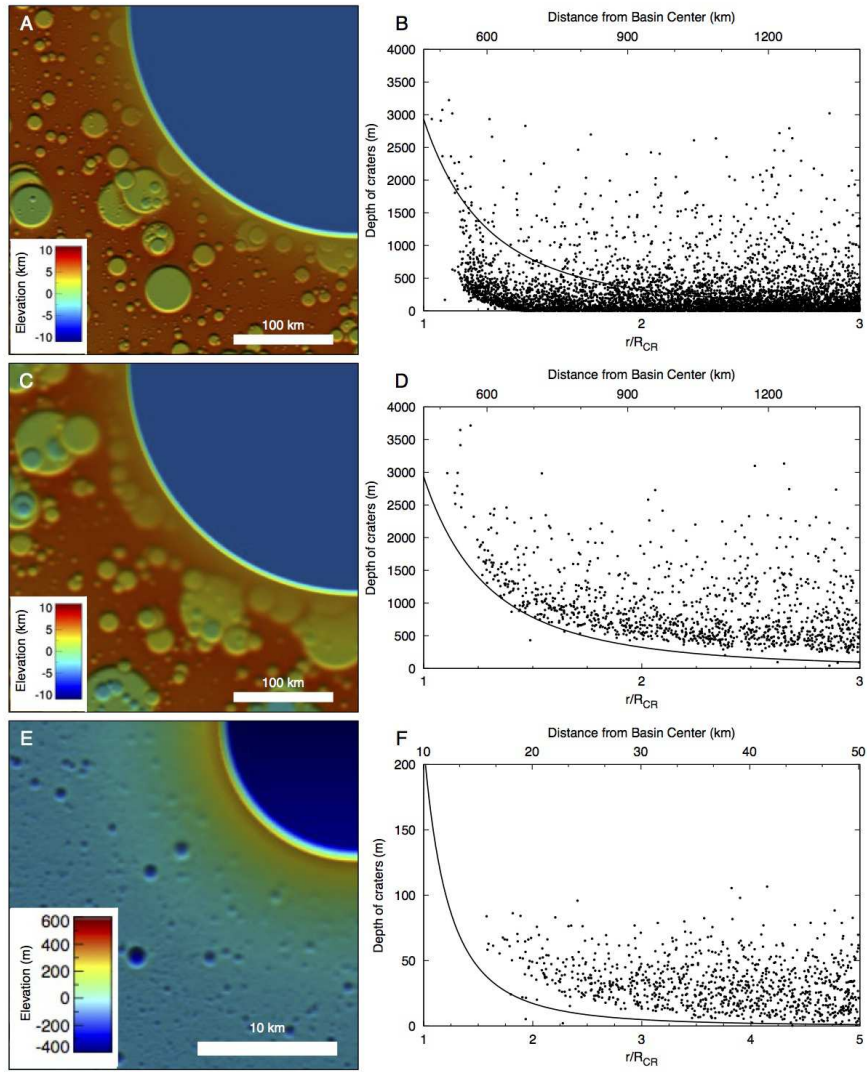} 
\caption{Examples of how {\em CTEM} handles crater erasure through ejecta burial. 
The left column shows a detail of a {\em CTEM}-generated DEM near a fresh crater. 
The right column shows depth of countable craters as a function of distance from the fresh crater center. 
A\&B) A $930\km$ diameter test crater (similar to Orientale) on a $2\km\px^{-1}$ grid. 
With topographic diffusion turned off, ejecta blanketing ``paints" the surface, preserving the topography of craters covered in ejecta. 
C\&D) The same crater as in A\&B, but this time with topographic diffusion turned on. 
Craters are now buried by ejecta; compare with Fig. 2a of \cite{Fassett:2011dj}.
E\&F) A $20\km$ diameter crater test crater on a $100\m\px^{-1}$ grid with the same diffusion coefficient used for the larger basin.
}
\label{f:soften-sequence}
\end{figure}

While the concept of ejecta burial is intuitive, modeling the process of burial by ejecta is not straightforward.
The reason why can be made clear with a series of figures.
\newtext{Fig.}~\ref{f:soften-sequence} shows shaded relief maps of a {\em CTEM} experiment designed to reproduce the Orientale crater.
In \newtext{Fig.}~\ref{f:soften-sequence}a, we show the result after a $930\km$ basin has been created at the center of the grid on a heavily cratered lunar terrain.
To generate this crater we used the TP2 (cold/strong upper mantle) basin scaling relationship derived by \cite{Potter:2012gm} to relate final rim diameter to transient crater diameter (see discussion in Section~\ref{subsec:basins}).
Few craters have been erased by the basin, as can be seen in \newtext{Fig.~\ref{f:soften-sequence}b}

\begin{figure}[htb] 
\centering
\includegraphics[width=\textwidth]{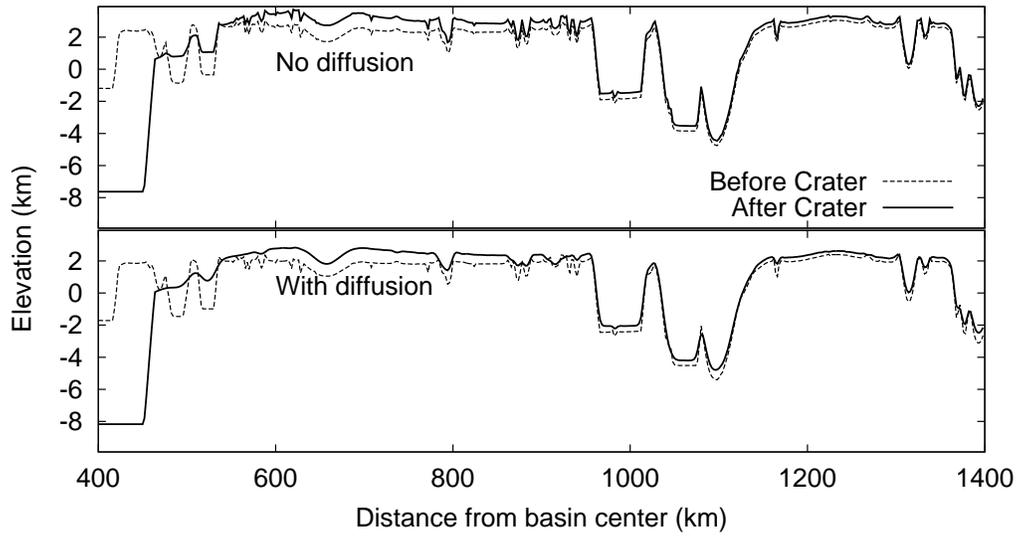} 
\caption{A) The pre-impact and post-impact surfaces are plotted for the simulation shown in Fig.~\ref{f:soften-sequence}a. Without the smoothing algorithm, {\em CTEM} ``paints'' the surface with ejecta, preserving the topography of the craters beneath. 
B) The same simulation shown in Fig.~\ref{f:soften-sequence}c, where we smooth the terrain by an amount proportional to the thickness of the ejecta blanket. 
In both plots, the rim of the crater is $465\km$ from the basin center.
The vertical profile has been exaggerated by $12.5\times$.}
\label{f:crosssection}
\end{figure}

The reason for the failure of the ejecta to erase craters can be seen in \newtext{Fig.}~\ref{f:crosssection}a, which plots the pre-impact and post-impact surface profiles along the centerline of the grid from the simulation shown in Fig.~\ref{f:soften-sequence}a. 
{\em CTEM} has effectively ``painted'' the surface with ejecta, preserving the topography of the craters beneath. 
This is clearly not a physically-plausible ejecta deposition model.
Real ejecta deposition is highly energetic, and the deposition process scours and transport material some distance away from the original ejecta impact point.

\newtext{
To model the ``blanketing'' effect of the ejecta blanket, we use topographic diffusion to smooth the terrain prior to emplacing ejecta. 
We use the diffusion equation of the form
\begin{equation}
\frac{\partial z}{\partial t} = \frac{\partial}{\partial x}\left(K_d\frac{\partial z}{\partial y}\right) + \frac{\partial}{\partial y}\left(K_d\frac{\partial z}{\partial y}\right),
\end{equation}
where $z$ is the elevation, $x$ and $y$ are the spatial dimensions on the surface, and $K_d$ is a diffusion coefficient, which can vary spatially. 
The diffusion model is discretized on our grid using a 2nd-order central-difference scheme.  

The diffusion coefficient, $K_d$, is a free parameter that is related to the local depth of ejecta.
\cite{Fassett:2011dj} used the depth of craters beyond the rim of Orientale to estimate the thickness of the Orientale proximal ejecta blanket.
By comparing the measured depth of a crater with its estimated original depth, \citeauthor{Fassett:2011dj} obtained an estimate of the ejecta infill as a function of distance from the rim of Orientale.
We used a similar technique in order to calibrate our ejecta blanketing model in {\em CTEM}.
We placed a $930\km$ diameter crater onto a heavily cratered surface on a $2\km\px^{-1}$ grid. 

Without our ejecta blanket diffusion model, craters beyond the rim of our test crater are preserved (Figs~\ref{f:soften-sequence}a \& b). 
We next fit a diffusion constant $K_d$ in our diffusion model such that craters are erased in a similar way as seen in the observational data of \cite{Fassett:2011dj}. 
The diffusion coefficient was found to be proportional to the ejecta blanket thickness (panels c \& d of Fig.~\ref{f:soften-sequence}). 
We then tested this model on smaller craters. 
When multiplied by the pixel size, the same diffusion constant found for the $930\km$ Orientale test also works for a $20\km$ crater on a $100\m\px^{-1}$ grid (panels e \& f of Fig.~\ref{f:soften-sequence}), a $2\km$ crater on a $10\m\px^{-1}$ grid and a $200\m$ crater on a $1\m\px^{-1}$ grid (not shown here). 
Our empirically determined diffusion coefficient is $K_d = 2.5\m^{-1}\s^{-1} b_e l_{px}$, where $b_e$ is the local ejecta thickness and $l_{px}$ is the length of a single pixel.
We note that each crater's ejecta profile is computed within {\em CTEM} using our excavation and ballistic transport model, and so they do not scale in a simple way with the crater size.
This gives us confidence that our model of crater erasure through proximal ejecta burial is robust.
}

\subsection{Crater counting calibration study}
\label{subsec:calibration}

Periodically {\em CTEM} performs a tally step to count all visible craters on the surface.
{\em CTEM}'s crater counting algorithm employs the use of identifying tags for craters that have formed on the surface, as described Section~\ref{subsubsec:cookie_cutting}.
This tag system has an advantage over algorithms that attempt to count craters on real surfaces: {\em CTEM} knows where all the craters formed.
This simplifies crater counting, in that {\em CTEM} does not need to employ any kind of feature-recognition algorithm in order to find potential craters. 
However, {\em CTEM} still needs to determine whether or not those craters that have escaped cookie-cutting are still features that a human would identify as a crater.
To accomplish this, we performed a calibration study with a human crater counter in order to determine what measurable qualities of {\em CTEM}-generated craterforms predicted their countability.

\newtext{
We based our calibration study on the classic experimental study described by \cite{Gault:1970bo}.
In that study, craters were produced in $2.5\m$ wide ``sandbox'' at the NASA Ames Research Center using projectiles and explosives to simulate heavily cratered lunar terrains.
\citeauthor{Gault:1970bo} formed the cratered surfaces using six projectiles that produced craters between $5\mm$ and $17\cm$, such that a crater was approximately twice as large as the next smallest.
The number of craters of a given size was constrained to be $k$ times as large as the next largest size class.
The terrains produced by the experiment were given names depending on the value of $k$: ``Terra Alta'' for $k=6$, ``Mare Exemplum'' for $k=10$, and  ``Mare Nostrum'' for $k=16$.
For the Terra Alta experiment, a seventh crater size of $35\cm$ (produced using explosives rather than a projectile) was added.
These crater size distributions may be approximately described by a cumulative power law given by $N_{>D}\propto D^{-p}$, where $p=\log\left(k+1\right)/\log 2$.

For our calibration study, we produced $5$ simulated surfaces designed to reproduced variations of the sandbox experiments of \cite{Gault:1970bo}.
Our simulations were constructed using either the ``Terra Alta'' ($k=6$) or ``Mare Nostrum'' surfaces ($k=16$), defined by \cite{Gault:1970bo}.
These simulated surfaces were designed to match as close as possible the experimental setup of \citeauthor{Gault:1970bo}.
For instance, we modeled the input size distribution as a step function, rather than a smooth power law as is normally done, to model the craters as discrete size classes.
Three of the simulations had a grid size of $2000^2$, one had a grid size of $500^2$, and one had a grid size of $100^2$.
The lowest resolution runs were performed so that we could obtain complete counts of very small craters on a given surface without overburdening our human counter.
We produced $10^4\m^{-2}$ craters on the Terra Alta surfaces and $10^5\m^{-2}$ on the Mare Nostrum surfaces.
We also included one high resolution Terra Alta simulation where we produced twice as many craters in order to obtain a highly saturated surface.

The collection of simulated surface DEMs were given to co-author Fassett, who was asked to count craters on each of the surfaces and report their diameters and center locations on the grid.
He identified a total of $744$ craters.
After receiving the list of identified craters for each simulated surface, we matched them with the list of craters produced by {\em CTEM} on that surface. 
The identified craters contained some error in size and position determination, and we had to adopt tolerance values for these two errors.
\cite{Robbins:2014kg} found that the typical position and diameter errors of professional crater counters was $5$--$10\%$.
We adopted a somewhat more generous position tolerance of $30\%$ relative to crater diameter and a diameter tolerance of $1/\sqrt{2}D_{true} < D_{reported} < \sqrt{2}D_{true}$, which is the width of a standard R-plot bin~\citepalias{CraterAnalysisTechniquesWorkingGroup:1979cs}.
This diameter tolerance is also larger than the factor of $2$ difference between the diameters of each crater size class, and therefore there is no ambiguity that a particular identified crater belongs to a particular diameter class.

Four craters were identified that did not correspond to any known crater produced by {\em CTEM}.
We dub these types of detections ``false positives,'' and because only a small number of these were reported, they are not expected to influence the overall results of our calibration and were discarded.
The calibration test results are reported in Table~\ref{t:calibration-test} lists the features of each of the simulation types, including the size class of the smallest reported craters, the number of craters produced that were larger than the smallest reported crater size, and the total number of craters that were counted.
}

\begin{figure}[htb]
\centering
\includegraphics[width=0.45\textwidth]{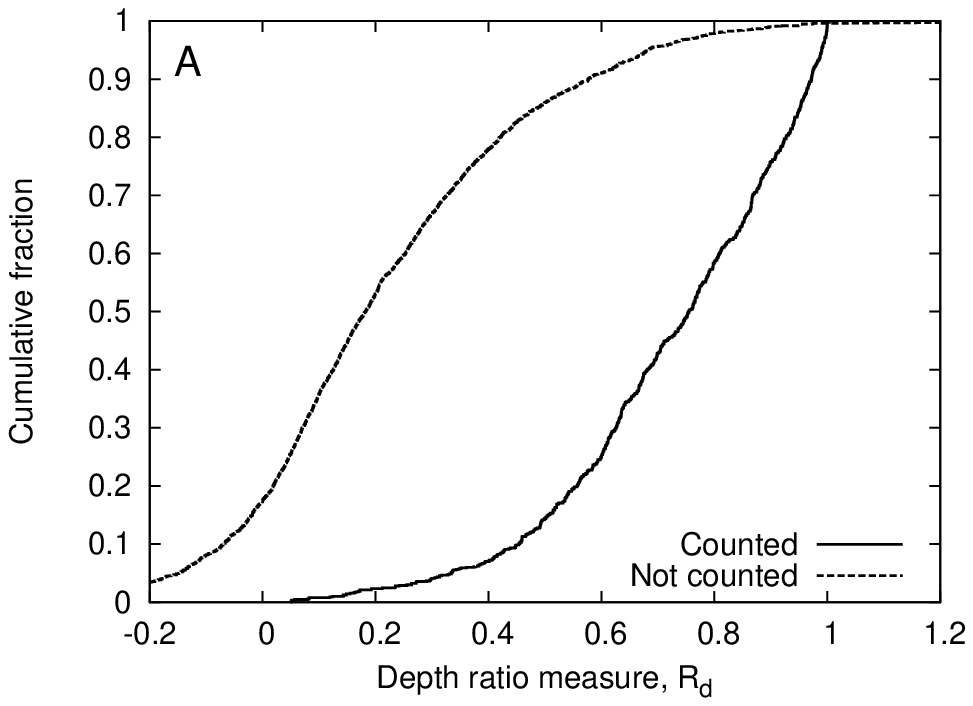}
\includegraphics[width=0.45\textwidth]{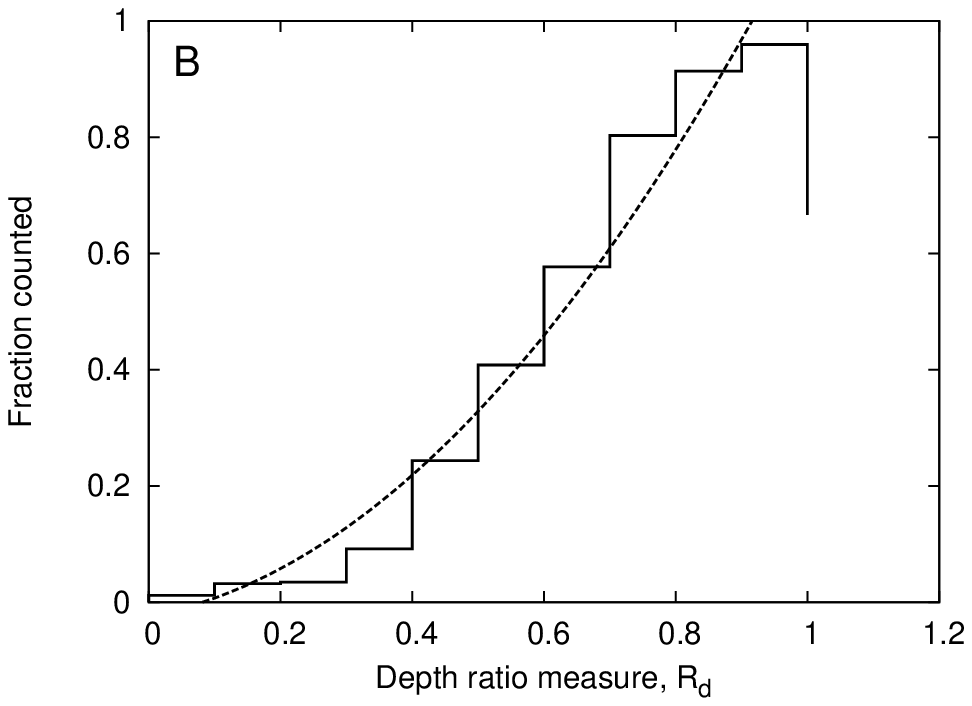}
\caption{ A) Cumulative fraction of craters with the depth ratio measure, ${R}_d$, below a given value. The counted craters are given as solid lines, and the uncounted craters are the dashed lines. 
B) Fraction of craters counted relative to total craters in bins of depth ratio measure, $R_d$. The bin width is $0.1$.  The dashed black line is a fitted curve to the complete crater set (black histogram), and is given by Equation~(\ref{e:pd}).
}
\label{f:depthratio}
\end{figure}

\newtext{
The goal of our analysis was to determine which measurable properties of a {\em CTEM}-generated crater correlated with countability.
} 
We identified two measures that both strongly correlate with the countability of a crater.
The first we call the depth ratio measure, ${R}_d$, which we define as:
\begin{equation}
R_d=\left<{h}_{rb}\right>/\left<{h}_{rb,0}\right>,
\label{e:Rddef}
\end{equation}
where $\left<{h}_{rb}\right>$ is the average rim-to-bowl height of the crater at the time of measurement, and $\left<{h}_{rb,0}\right>$ is the original average rim-to-bowl height.
\newtext{
These measurements are made relative to a reference plane that is determined by the average slope of the terrain just prior to crater emplacement.
}
This allows us to accurately measure rim-to-bowl heights on craters that form on slopes (such as small craters that form on the inner walls of larger craters).
The cumulative fraction of craters below a given $R_d$ is plotted in \newtext{Fig.}~\ref{f:depthratio}a.
We have separated out the subset of $111$ test craters as red lines in this figure, showing that there was no substantial difference between the test craters and the complete set.

\newtext{Fig.}~\ref{f:depthratio}a shows that there is a strong correlation between $R_d$ and the countability of a crater. 
This correlation can further be quantified by placing craters in bins of $R_d$ value, and then calculating the fraction of craters in each bin that were counted.
We plot the fraction of craters counted in bins of $R_d$, with a bin width of $0.1$, in \newtext{Fig.}~\ref{f:depthratio}b. 
We then perform a least squares fit to the data to produce a model of the probability of being counted as a function of $R_d$, which we call the depth ratio probability function $p_d$.
We found that a second order polynomial was a better fit than a linear function. 
The best fit functional form of the depth ratio probability function is given as equation~(\ref{e:pd}).
\newtext{
\begin{equation}
p_d(R_d)=
\begin{cases}
	0 &: R_d \leq 0.082\\
	-0.024 + 0.209 R_d + 0.993R_d^2 &: 0.082 < R_d < 0.92\\
	1 &: R_d \geq 0.92
\end{cases}
\label{e:pd}
\end{equation}
}
The value of $p_d$ for any given crater is a probability that a crater with its corresponding $R_d$ value will be counted by a human or not.

\begin{figure}[htb]
\centering
\includegraphics[width=0.45\textwidth]{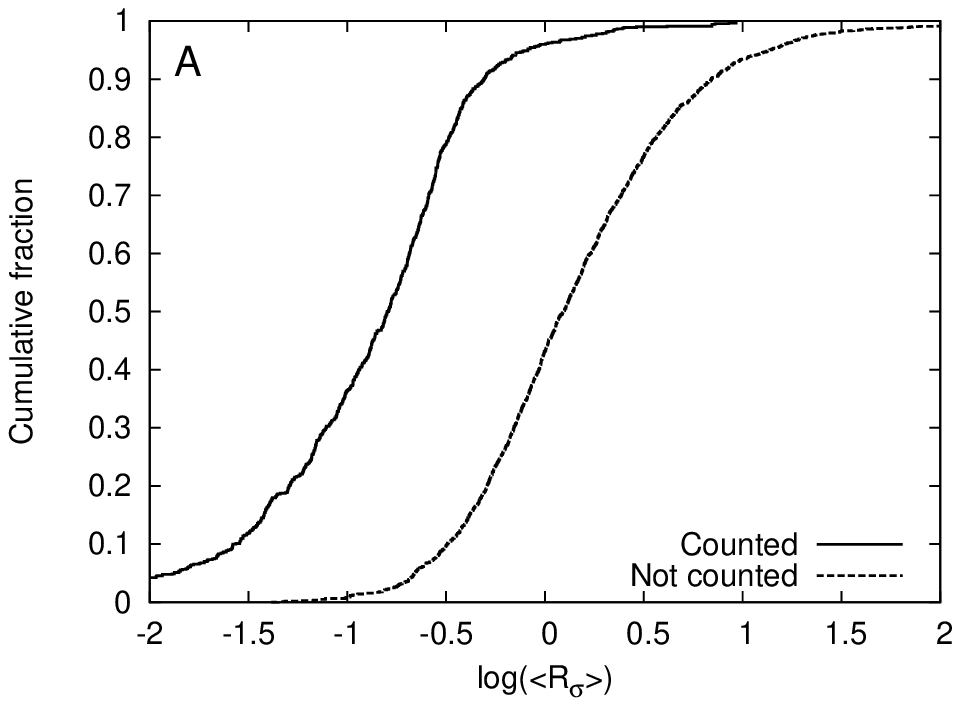}
\includegraphics[width=0.45\textwidth]{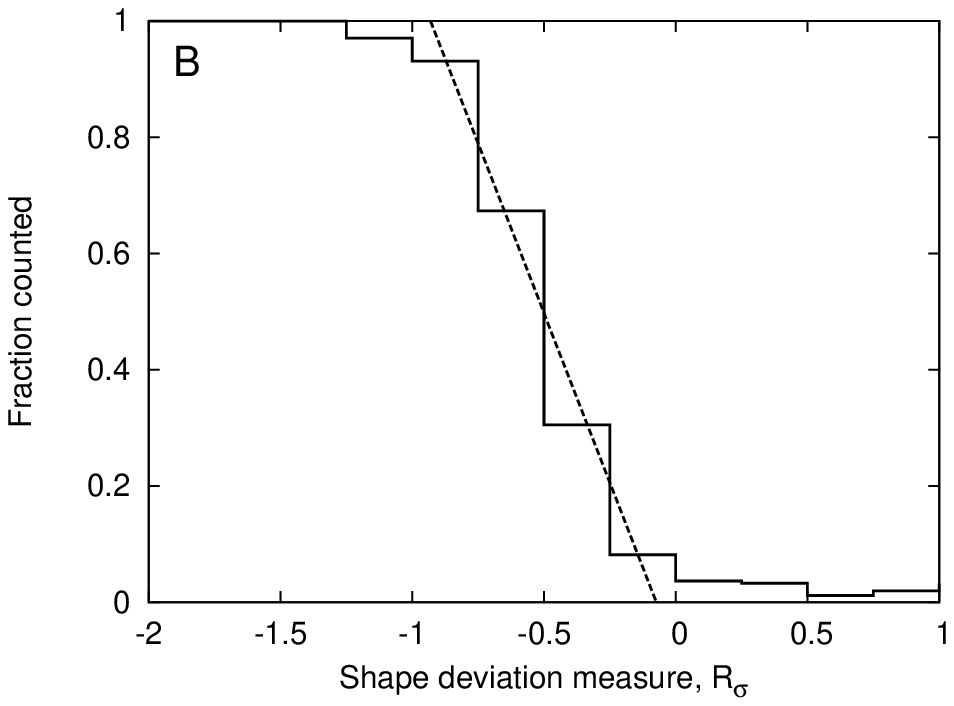}
\caption{A) Cumulative fraction of with the shape deviation measure, ${R}_\sigma$, below a given value. The counted craters are given as solid lines, and the uncounted craters are the dashed lines. 
B) Fraction of craters counted relative to total craters in bins of shape deviation  measure, $R_\sigma$. The dashed black line is a fitted curve to the complete crater set (black histogram), and is given by Equation~(\ref{e:psig}).}
\label{f:sigmadev}
\end{figure}

We also identified a second measure that also correlates strongly with the countability of the craters.
The second measure we call the shape deviation measure, $R_\sigma$, which we define as:
\newtext{\begin{equation}
{R}_\sigma = \log\left(\sigma/\left<{h}_{rb}\right>\right),
\label{e:Rsigdef}
\end{equation}}
Here, $\sigma$ is the standard deviation of the average difference between the elevation of the crater pixels at the time of measurement and its original elevation after formation.
\newtext{Again, measurements are made relative to a reference plane that is at the average terrain slope prior to crater emplacement.}
For any given crater that occupies $N$ pixels, $\sigma$ is defined as:
\begin{equation}
	\sigma = \sqrt{\frac{1}{N}\Sigma_{i=1}^N\left[\left(h_i - h_{i,0}\right) - \mu\right]^2},
	\label{e:sigmadef}
\end{equation}
where $h_i$ is the elevation of pixel $i$ and $h_{i,0}$ is the elevation that pixel $i$ had just after the crater formed, and
\begin{equation}
	\mu = \frac{1}{N}\Sigma_{i=1}^N\left(h_{i} - h_{i,0}\right).
	\label{e:mudef}
\end{equation}
The cumulative fraction of craters below a given $R_\sigma$ is plotted in \newtext{Fig.}~\ref{f:sigmadev}a.

\newtext{Fig.}~\ref{f:sigmadev}a shows that there is a strong correlation between $R_\sigma$ and the countability of a crater. 
We can perform a similar analysis that we did for $R_d$ by placing craters in bins of $R_\sigma$ value, and then calculating the fraction of craters in each bin that were counted.
We plot the fraction of craters counted in bins of $R_\sigma$, with a bin width of $0.25$, in \newtext{Fig.}~\ref{f:sigmadev}b. 
We then perform a least squares fit to the data to produce a model of the probability of being counted as a function of $R_d$, which we call the depth ratio probability function $p_d$.
We found that a linear function fit the data well. 
The best fit functional form of the shape deviation probability function is given as equation~(\ref{e:psig}).
\newtext{
\begin{equation}
p_\sigma(R_\sigma)=
\begin{cases}
	1 &: R_\sigma \leq -0.93\\
	-0.0856-1.167R_\sigma &: -1.5 < R_\sigma < 0.05\\
	0 &: R_\sigma \geq -0.073
\end{cases}
\label{e:psig}
\end{equation}
}

\begin{figure}[htb]
\centering
\includegraphics{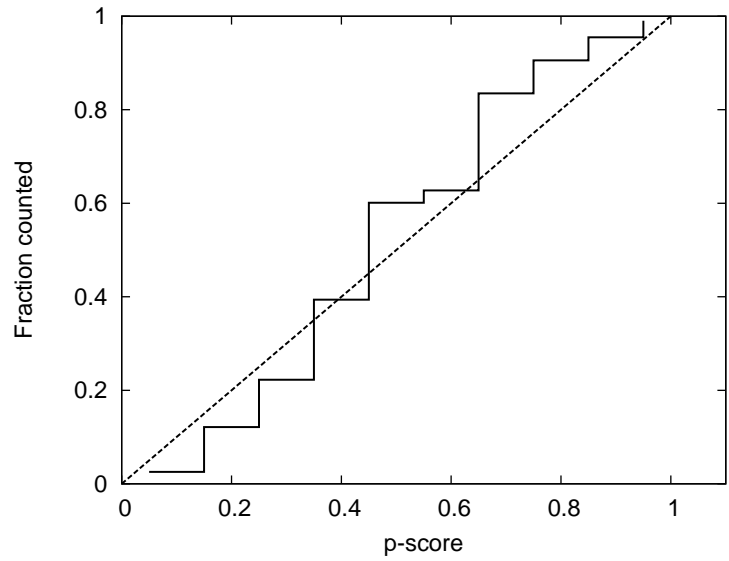}
\caption{Fraction of craters counted relative to total craters in bins of p-score, which is the smallest of either $p_d$, given by equation~(\ref{e:pd}) or $p_\sigma$, given by equation~(\ref{e:psig}). The black histogram is for all craters in the calibration set. The dashed black line is y(x) = x, and it is the expected result if the p-score was exactly the probability of counting.
}
\label{f:scorehist}
\end{figure}

The two measures that we identified, $R_d$ and $R_\sigma$, are correlated.
Therefore we cannot treat them as independent probabilities.
However, we have found that taking the minimum value of either $p_d$ or $p_\sigma$, which we dub the ``p-score,''  predicts the probability that any given crater is countable better than either measure alone.
A histogram of the fraction of craters that were counted in bins of p-score (for all $560$ craters in our calibration set) is plotted  in \newtext{Fig.}~\ref{f:scorehist}.
The dashed line is the $y(x) = x$ line, and would be the ideal case where the p-score exactly predicts the probability of any crater being counted.
Our calibration parameters match the probability of counting quite well, closely adhering to the $y(x)=x$ line.

We implement our calibration results into {\em CTEM} using the p-score. 
For every crater, we evaluate its p-score using by first calculating $R_d$ and $R_\sigma$ using equations~(\ref{e:Rddef}) and (\ref{e:Rsigdef}), respectively. 
Then we calculate $p_d$ and $p_\sigma$ using equations~(\ref{e:pd}) and (\ref{e:psig}), respectively.
The minimum of the two becomes the p-score.
\newtext{Ideally, the p-score would be used as a probability that the crater is counted or not. 
Implementing the p-score as a pure probability function yielded complications to the code, because low-p craters could have their p-scores artificially inflated if a new crater happened to form with a similar size and position as an old, degraded crater.
This results in an artificial secular increase in the number of craters on a heavily cratered terrain over time.
In order to prevent this, we use a p-score of $0.5$ as a threshold for countability.}
\newtext{If the p-score is less than $0.5$,} then the crater is not counted, and the record of the crater's existence is obliterated from the grid so that it cannot be later confused for another similar-sized crater that might later happen to form close to the old crater's location.

\begin{figure}[htb]
\centering
\includegraphics[width=\textwidth]{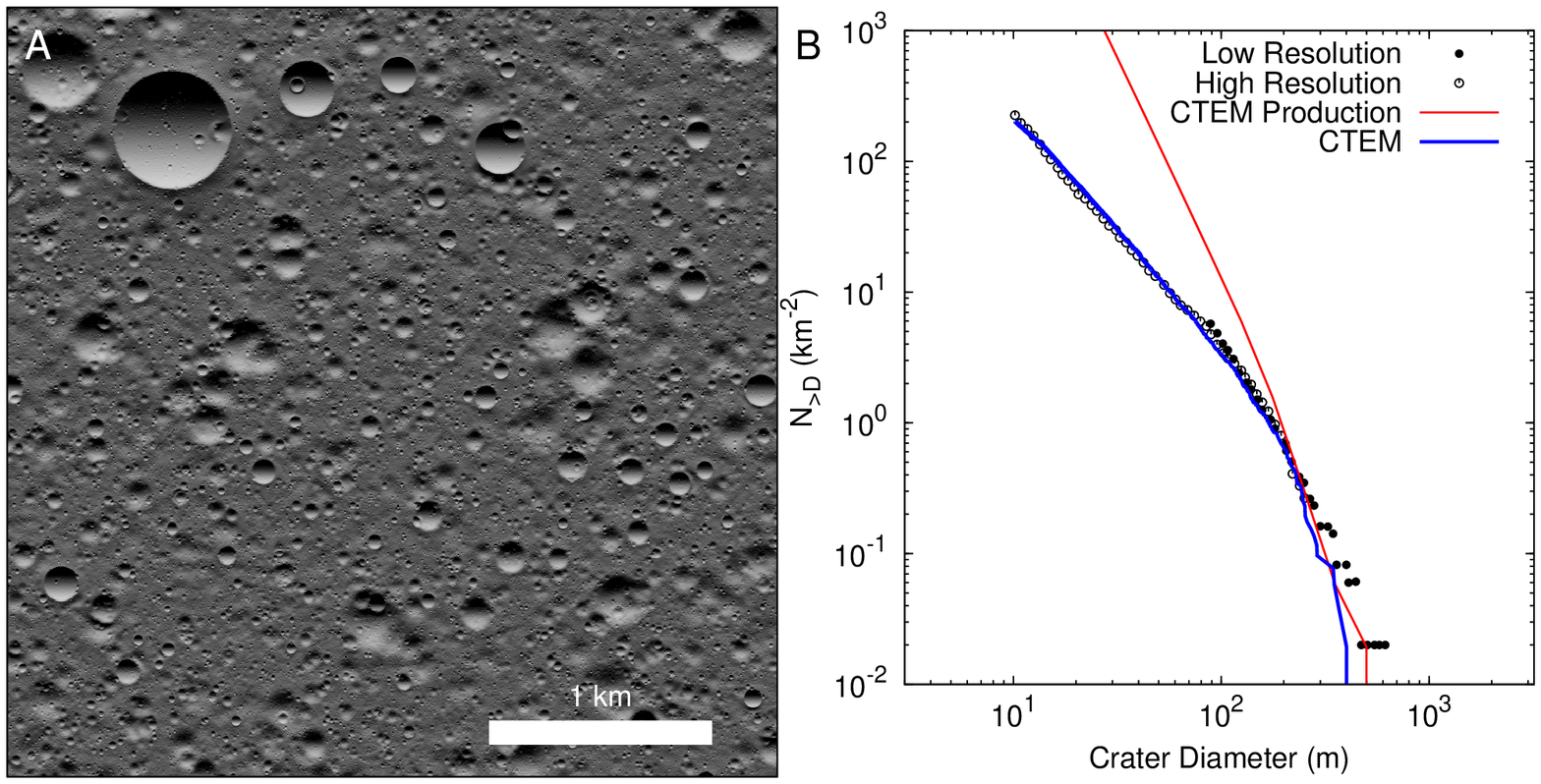}
\caption{In this simulation {\em CTEM} is used to model Sinus Medii, a small mare deposit at the sub-Earth point of the Moon that was emplaced between $3.63$--$3.79\Gy$ ago~\citep{Hiesinger:2010ke}. 
This mare is often used as a case study in crater equilibrium (or saturation equilibrium), due to the ``break'' in the power law slope of the crater SFD at $\sim100\m$ crater diameter~\citep{Gault:1970bo,Marchi:2012hm}. 
The left-hand panel shows a shaded reproduction of a segment of the full $2000\times2000\px$ DEM surface ($3.6\m\px^{-1}$ resolution). 
The right-hand panel shows a cumulative size-frequency distribution of {\em CTEM} generated craters (red), the countable craters (blue), and observational crater counts of Sinus Medii from \cite{Gault:1970bo}  (black circles). 
An impactor population with of the form $N_i\propto D_i^{-3.25}$ was used to generated the craters, and a ``dry soil'' model for the regolith material properties was assumed for the crater scaling relationship was used where $\mu=0.41$, $K_1=0.24$, and $\bar{Y}=0.18\MPa$~\citep{Holsapple:1993js}.
}
\label{f:SinusMedii}
\end{figure}

\newtext{
We verified our calibrated counting using the ``classic'' example of a terrain in cratering equilibrium: Sinus Medii.
Sinus Medii is a small mare deposit at the sub-Earth point of the Moon that was emplaced between $3.63$--$3.79\Gy$ ago~\citep{Hiesinger:2010ke}. 
This mare is often used as a case study in crater equilibrium (or saturation equilibrium), due to the ``break'' in the power law slope of the crater SFD at $\sim100\m$ crater diameter~\citep{Gault:1970bo,Marchi:2012hm}. 
We performed a test in {\em CTEM} designed to reproduce the cratering history of Sinus Medii. 
An impactor population with of the form $N_i\propto D_i^{-3.25}$ was used to generated the craters, and a ``dry soil'' model for the regolith material properties was assumed for the crater scaling relationship was used where $\mu=0.41$, $K_1=0.24$, and $\bar{Y}=0.18\MPa$~\citep{Holsapple:1993js}.

\newtext{Fig.}~\ref{f:SinusMedii} shows the results of our simulation.
\newtext{Fig.}~\ref{f:SinusMedii}a shows a shaded reproduction of a segment of the full $2000\times2000\px$ DEM surface ($3.6\m\px^{-1}$ resolution). 
Multiple generations of ejecta blanketing and the resultant topographic diffusion give the surface a ``soft'' appearance, consistent with the appearance of real lunar terrains at this scale.
\newtext{Fig.}~\ref{f:SinusMedii}b shows a cumulative size-frequency distribution of {\em CTEM} generated craters (red), the countable craters (blue), and observational crater counts of Sinus Medii from \cite{Gault:1970bo}  (black circles). 
With our calibrated crater count algorithm we are able to achieve a very good fit to the observed crater counts of this mare terrain at this scale.
We also do not see a secular increase in the small-crater abundance once it reaches the equilibrium line.
The mismatch at the largest crater size is a resolution effect due to the limited size of our grid.
}

\subsection{Impactor population} 
\label{subsec:impactor_population}
In this work we wish to test whether the highlands impacting population originated in the main asteroid belt, as suggested by~\cite{Strom:2005fo}. 
This  constrains the impactor size frequency distribution, impactor density, and impact velocity distribution.
We use the observed main asteroid belt size frequency distribution (SFD) for our impactors.
For $D_{i}>4\km$ the SFD of the main asteroid belt has been well characterized by the Wide-field Infrared Survey Explorer (WISE) spacecraft.

The WISE spacecraft detected over 100,000 asteroids in thermal infrared wavelengths, and used published estimates of optical brightness to constrain individual asteroid albedos, and therefore estimate diameters~\citep{Mainzer:2012de,Masiero:2011jc}.
We use the sizes of main belt asteroids from the Pass 1 WISE/NEOWISE data as reported by \cite{Masiero:2011jc}.
Because the WISE survey was incomplete for \newtext{$D_{i}<5\km$}, and the Pass 1 data is not debiased, we created a model main belt SFD that used WISE data for \newtext{$D_{i}>5\km$ and a fit to the Sub-Kilometer Asteroid Diameter Survey (SKADS) results published in \cite{Gladman:2009cx} for $D_i<5\km$}.

Our impact velocity distribution shown in \newtext{Fig.}~\ref{f:cehistory-impacts-Moon} is derived from an N-body simulation of the dynamical diffusion of main belt asteroids into the NEA region~\citep{Minton:2010cn,Yue:2013bx}. 
The RMS velocity of this distribution is $18.3\km\s^{-1}$.
We adopted $2.5\gm\cm^{-3}$ as the density of our impactors in line with typical densities of S-type asteroids~\citep{Britt:2002wx}, assuming that most lunar impactors derive from the S-dominated inner main asteroid belt~\citep{Bottke:2006ge}.

For each impact, {\em CTEM} draws an impactor size and velocity from the input distributions and determines a random location on the surface to place the impactor.
The code also selects a random impact angle based upon the canonical formula whose derivation is summarized in \cite{Pierazzo:2000bx}: 
\begin{equation}
\label{obliqueimpact}
dP = 2 \sin \theta \cos \theta d \theta  \>.
\end{equation}
The normalized, integrated (cumulative) form of this equation is simply:
\begin{equation}
\label{obliqueimpact_cumulative}
P = \sin^2 \theta   \>,
\end{equation}
where Eq.~\ref{obliqueimpact_cumulative} is inverted to generated a random impact angle $\theta$ as a function of a randomly generated probability ($P$ between 0 and 1) for each simulated impact in the model.

The effects of our velocity and angle distributions on the scaling relationship is shown in Fig.~\ref{f:scalecontour}.
Here we plot the average number of craters produced in $500$ {\em CTEM} simulations as a density map. 
The log of the average number of craters within a bin with a size of $0.1\times0.1$ log diameter (km) is plotted as the contour value.
This also includes the effects of our basin scaling, which we use for $D_i>35\km$.
Because the scaling parameters depend differently on size (mass) and velocity, there is a size dependence in how the various distributions shape the distribution of impactor sizes for a given crater size.

\section{Lunar cratering simulations}
\label{sec:global}
Our goal is to test the hypothesis that impactors with the main belt asteroid SFD can produce the observed lunar highlands crater SFD.
Our comparison data set is the catalog of all observed lunar craters with $D > 20\km$ obtained using the Lunar Orbiter Laser Altimeter (LOLA) aboard the Lunar Reconnaissance Orbiter spacecraft~\citep{Head:2010fy}.
We studied this problem in two steps.
First we performed a series of regional simulations designed to determine the minimum level of cratering (i.e.~impactor flux exposure) needed to reproduced the observed abundance of craters found on the lunar highlands.
\newtext{Next we performed a series of global simulations designed to determine whether or not the best fit level of cratering obtained in the regional simulations is consistent with the observed number of large lunar basins.
The regional constraint is a lower limit on the abundance of mid-sized craters, while the global constraint is an upper limit on the number of large basins.
Because of the nature of the Monte Carlo technique, the ``best fit'' for any given constraint is 50\%.
That is, half the runs for a given level of cratering satisfy the constraint, while half do not.
}
\newtext{The total amount of cratering on a given run is parameterized by the quantity of expected impactors larger than $10\km$ in diameter, $\bar{N}_{pf}(D_i>10\km)$, per unit area.
The actual number of impactors, $N_i(D_i>10\km)$, will vary from run to run due to the nature of the Monte Carlo method, but over many runs the average number will equal the expected number.
}
We step through values of \newtext{$\bar{N}_{pf}(D_i>10\km)$} until we match the observed abundance of craters on the lunar highlands from our crater catalog.
We make use of the relative crater density, or R-value, which is defined by \citepalias{CraterAnalysisTechniquesWorkingGroup:1979cs} as:
\newtext{
\begin{equation}
	R = \frac{\bar{D_c}^3N_c}{A\left(b_2-b_1\right)}.
	\label{e:Rdef}
\end{equation}
}
where \newtext{$N_c$} is the number of craters within bins with boundaries $\left(b_1,b_2\right)$.
We use standard R-plot bins where the bin boundaries are given as $1\km\times\left(2\right)^{n/2}$, and $n$ is an integer~\citepalias{CraterAnalysisTechniquesWorkingGroup:1979cs}.
The geometric mean diameter, \newtext{$\bar{D}_c$}, of the bin is defined as:
\newtext{
\begin{equation}
	\bar{D}_c=\left(\Pi_{j=1}^ND_{c,j}\right)^{1/N}
	\label{e:geomD}
\end{equation}
}
where \newtext{$D_{c,j}$} are the individual craters in the bin.
Alternatively, 
\newtext{
\begin{equation}
	\bar{D}_c=\sqrt{b_1b_2}
	\label{e:geomD2}
\end{equation}
}
is used when the individual crater diameters are not available.

We will also make use of R-plots, which are plots of the log of $R$ as a function of the log of \newtext{$D_c$}, as a way of comparing size distributions of observed craters and modeled craters.
\newtext{For the regional simulation step, we use only the subset of craters in the LOLA catalog that occur on the lunar highlands, excluding Orientale and SPA basins.
This region covers an area of $2\times10^7\km^2$ of the Moon, and is shown in Fig.~\ref{f:nonmarenoSPA}} 
In particular, craters in the size range $90.5$--$128\km$ make for a very useful diagnostic for determining how many impacts are required to match the regional lunar highlands cratering record, because the relative crater density, or R-value, of lunar highlands craters in this size range is at a peak. 

Once we have determined the probability that a given value of \newtext{$\bar{N}_{pf}(D_i>10\km)$} will reproduce the observed crater density, we next performed a similar set of runs for a global lunar surface.
In these runs, we define a basin constraint based on the observed abundances of lunar basins in our LOLA crater catalog, where we adopt the definition that a basin is any crater with \newtext{$D_c>300\km$}.
The basins are observable, even if they are nearly completely filled with mare, such is the case with Imbrium.
Recently measurements of the gravity signature of basins on the Moon obtained from the GRAIL spacecraft have revealed that the observed number of basins is a largely complete inventory of all basins that have formed on the Moon since the crust solidified~\citep{Neumann:2013wp}.
In particular, the largest impact structure on the Moon is the $\sim2500\km$ South Pole-Aitken basin (SPA), which is a very prominent and deep topographic feature, and acts as a strong constraint on the total impact history of the Moon since the crust solidified.
SPA appears to pre-date all other features on the Moon~\citep{Wilhelms:1987vb}, so it is somewhat ambiguous whether its formation is linked to the rest of the lunar basin formation process.
Nevertheless, it is unlikely that any other basin as large or larger than SPA postdates its formation, and its nearest rival is the $\sim1160\km$ Imbrium basin.
Because SPA apparently pre-dates the entirety of the lunar highlands cratering record, we would be well justified in ignoring it in our study.
Instead, we will adopt the conservative assumption that the SPA impactor originated from the same population that the rest of the highlands did, but by chance impacted near the beginning of the observable lunar cratering record. 
\newtext{For our simulations, we adopt the constraint that we must produce no more $1$ basin with $D_c>1200\km$ (the size of Imbrium).
For the E-belt simulations our basin constraint is that we must produce no basins with $D_c>1200\km$.}

For each set of conditions we performed $100$--$1000$ {\em CTEM} simulations of the lunar surface.
We tally the countable craters in each simulation, and bin them.
However, we plot our results in a way that is somewhat unusual for crater counting.
Usually crater counts are reported using error bars that are scaled by $\pm{N}^{1/2}$ in size to reflect the assumption that the variability in the number of craters for a given amount of flux is Poisson-distributed~\citepalias{CraterAnalysisTechniquesWorkingGroup:1979cs}. 
However, we make no a priori assumption as to how crater variability is characterized.
While \newtext{{\em impact}} events are well described with Poisson statistics, because the area of a cratered surface is finite and craters may obliterate one another, {\em craters} are not strictly Poisson-distributed.
Large craters are more effective obliteration agents than small craters, and as craters approach a size comparable to the counting area, Poisson statistics are a poor model for the intrinsic variability of craters.
By performing a large number of runs, we can directly estimate the variability of crater counts directly without having to assume that they are Poisson-distributed.

To report our {\em CTEM}-derived results for multiple runs with the identical parameters (except for the random number generator seed), we make use of ``box and whisker plots.''
This style of plot contains a box within an error bar.
The box is drawn to span the $25\%$ of data points above the median and the $25\%$ of data points below the median. 
The value of the median is drawn as a horizontal line within the box.
The error bars enclose $99\%$ of the data.
Because we report our model statistics, we have no need to estimate statistical variations in the observed data set.
Therefore our observational data set (the catalog of LOLA crater of \newtext{$D_c>20\km$}) is plotted simply as points with no error bars.
An example of data plotted in this fashion is given in \newtext{Fig.}~\ref{f:regional-boxplot}.

\subsection{Regional runs with the $N_{90.5-128\km}$ constraint.}
\label{subsec:regional}
Our regional {\em CTEM} simulations were performed on a $2000\times2000$ pixel grid at a resolution of $2.24\km\px^{-1}$.
Our catalog of highlands craters includes craters on an area of $2\times10^7\km^2$ of the lunar surface that excludes SPA and Orientale \newtext{(see Fig.~\ref{f:nonmarenoSPA})}.
This catalog contains no craters larger than $628\km$ in diameter.
In these simulations we deliberately restricted our cratering model to produce only craters less than this maximum size. 
The craters in this region were not greatly effected by mare emplacement, and therefore more closely reflect the crater density of the pre-mare lunar surface than the global catalog.
We bombarded a simulated lunar surface with an amount of craters equal to \newtext{$\bar{N}_{pf}(D_i>10\km)=3.5\times10^{-6}\km^{-2}$--$6\times10^{-6}\km^{-2}$}.

We found that the bin that spans \newtext{$D_c=90.5$}--$128\km$ to be a useful diagnostic of the success of cratering of a given terrain, as it has the highest value of $R$ in the regional data set.
The total number of observed craters in LOLA lunar highlands regional data for this bin is 198, and we set as our constraint that our simulated surface must also have this many craters in this size range to be considered a success.
\newtext{Fig.}~\ref{f:regional-boxplot} shows results of four sets of {\em CTEM} simulations in R-plot form.
\newtext{For $\bar{N}_{pf}(D_i>10\km)<4.0\times 10^{-6}\km^{-2}$}, none of the runs produced the observed number of craters in the $90.5$--$128\km$ bin.
\newtext{For $\bar{N}_{pf}(D_i>10\km)=5.0\times 10^{-6}\km^{-2}$, $55.6\pm2.4\%$ of runs satisfied the regional mid-sized crater constraint, and therefore is close to the best fit cratering abundance for the main belt size distribution.}
We plot these statistical results as red points in \newtext{Fig.}~\ref{f:combined_constraint}.

\subsection{Global lunar surface runs with the basin constraint}
\label{subsec:global}
Our global {\em CTEM} simulations were performed on a $2000\times2000$ pixel grid at a resolution of $3\km\px^{-1}$.
We bombarded a simulated global lunar surface with an amount of craters equal to \newtext{$\bar{N}_{pf}(D_i>10\km)=5\times10^{-7}\km^{-2}$--$6\times10^{-6}\km^{-2}$}.
\newtext{We set a constraint that there should be no more than 1 basin with $D_c>1200\km$ and ended the run if a crater this large was generated.}
\newtext{Fig.}~\ref{f:combined_constraint} shows the fraction of simulations that fit our basin constraints as a function of \newtext{$\bar{N}_{pf}(D_i>10\km)$}.
As we saw earlier only runs with \newtext{$\bar{N}_{pf}(D_i>10\km)>4\times10^{-6}\km^{-2}$} satisfied the regional constraint\newtext{, with $5\times10^{-6}\km^{-2}$ being close to a best fit value}.
However, for \newtext{$\bar{N}_{pf}(D_i>10\km)=4\times10^{-6}\km^{-2}$}, only \newtext{$0.8\pm0.28\%$} of runs satisfied the basin constraint.
\newtext{Because we end the runs at the moment that a crater larger than $1200\km$ in diameter is generated, we can determine the mean value of $\bar{N}_{pf}(D_i>10\km)$ at the time that the first constraint-violating basin formed.
We find that the global constraint is violated on average at $\bar{N}_{pf}(D_i>10\km)=9.2\times10^{-7}\km^{-2}$.
This is approximately a factor of $5$ less cratering than the best fit value of $\bar{N}_{pf}$ determined using the regional constraint. 
}

We then compared the average global cratering record for those runs that satisfied the basin constraints and compared the crater size distribution of smaller craters. 
Our two constraints were only \newtext{very rarely satisfied for $\bar{N}_{pf}(D_i>10\km)=5\times10^{-6}\km^{-2}$, where $8$ out of $1000$} global simulations satisfied the basin constraint. 
We plot the resulting crater R-plot distributions for the runs with \newtext{$\bar{N}_{pf}(D_i>10\km)=1\times10^{-6}\km^{-2}$ and $\bar{N}_{pf}(D_i>10\km)=5\times10^{-6}\km^{-2}$} in \newtext{Fig.}~\ref{f:basincon-boxplot}. 
\newtext{Although $\bar{N}_{pf}(D_i>10\km)=1\times10^{-6}\km^{-2}$ is very near the best fit value of cratering that satisfies the basin constraint, none of the runs had enough craters to satisfy the regional constraint.}
As we showed in our regional simulations in Section~\ref{subsec:regional}, we needed \newtext{$\bar{N}_{pf}(D_i>10\km)>4\times10^{-6}\km^{-2}$} in order to match the observed abundance of craters in the $90.5\km$--$128\km$ size range.
The global constraint, by contrast, is better matched at \newtext{$\bar{N}_{pf}(D_i>10\km)=9.2\times10^{-7}\km^{-2}$}.

\subsection{Runs that test the E-belt hypothesis}
\label{subsec:ebelt}
The region inward of $\sim2\AU$ is currently unstable due to the presence of the $\nu_6$ secular resonance~\citep{Williams:1981hm}.
If the giant planets formed in a more compact configuration, the $\nu_6$ would have been farther from the Sun than its present location~\citep{Minton:2011ir,Agnor:2012bo}, and the region between the inner main asteroid belt and Mars could have been filled with asteroids.
Under the E-belt hypothesis, described in \citep{Bottke:2012ce}, the arrival of the $\nu_6$ resonance after giant planet migration would have been responsible for destabilizing this innermost portion of the asteroid belt (the E-belt), and the impactors associated with the LHB would have primarily come from this region.
The E-belt, plus a small contribution from the main belt, could only supply enough large impactors to produce the sequence of basins beginning with Nectaris.
This population was assumed to have had a similar size-frequency distribution as the current inner main asteroid belt, however the impact velocity of this population was somewhat higher, with a median velocity of $21\km\s^{-1}$ instead of the $18.3\km\s^{-1}$ that we used in the previous sections.

Based on LOLA crater counts, crater density on Nectaris basin is N$(64)=17\pm5$ and N$(20)=135\pm14$~\citep{Fassett:2012hr}.
Here N$(D)$ refers to the \newtext{$N_{>D_c}$} per $10^6\km^2$ surface area.  
\cite{Fassett:2012hr} also report $14$ basins younger than Nectaris, plus Nectaris itself.
The largest crater younger then Nectaris is the \newtext{$D_c\simeq1200\km$} Imbrium basin.
We performed a series of regional runs similar to those described in Section~\ref{subsec:regional}, as well as a series of global runs similar to those described in Section~\ref{subsec:global}.
The results are shown in \newtext{Fig.}~\ref{f:ebelt_constraint}.
The fraction runs for a given value of \newtext{$\bar{N}_{pf}(D_i>10\km)$} that produced N$(20)$ and N$(64)$ densities within the range determined by \cite{Fassett:2012hr} for Nectaris are plotted as green triangles and red squares respectively.
\newtext{For clarification, because here we are using crater densities as our observational constraint, rather than total number of craters as in Sections~\ref{subsec:regional} and \ref{subsec:global}, we do not use $50\%$ match as our criteria for ``best fit.'' 
Here we simply use the reported uncertainty ranges of the crater densities to determine whether model runs at a given number of impactors are a good fit or not.}

\newtext{The N$(64)$ densities suggest values of $\bar{N}_{pf}(D_i>10\km)=1.5$--$2.25\times10^{-6}\km^{-2}$ in order to reach the observed crater densities on Nectaris.
The N$(20)$ densities suggest somewhat higher values of $\bar{N}_{pf}(D_i>10\km)=2.25$--$2.75\times10^{-6}\km^{-2}$, however this could simply imply that the dataset is either overabundant in $\sim20\km$ craters, or the slope our SFD for craters with $D_c<100\km$ is too shallow. 
To avoid any ambiguity in the $D_c\simeq 20\km$ crater counts we adopt the N$(64)$ densities as our constraint.
}
\newtext{For our E-belt basin constraint, we require that runs produce no basins with $D_c>1200\km$.} 
The fraction of runs for a given value of \newtext{$\bar{N}_{pf}(D_i>10\km)$} that satisfied the basin constraint is plotted as black circles in Fig.~\ref{f:ebelt_constraint}. 
\newtext{
Only $16\pm2.3\%$ of runs at $\bar{N}_{pf}(D_i>10\km)=1.75\times10^{-6}\km^{-2}$ satisfied the basin constraint, and $0.59\pm0.8\%$ of runs satisfied it at $\bar{N}_{pf}(D_i>10\km)=2.25\times10^{-6}\km^{-2}$.
} 

\section{Discussion \newtext{\& Conclusions}}
\label{sec:discussion}
We defined two constraints that must be satisfied simultaneously in order for the main asteroid belt size-frequency distribution to produced the observed lunar highlands crater size-frequency distribution. 
Our two constraints are a regional one (we must produce at least the total number of observed \newtext{$D_c\simeq100\km$} craters on the highlands), and a global one (we must produce no more than the observed number of basins and none larger than the largest observed basin). 
Our results indicate a very low probability in matching both of these constraints with an impactor population resembling the modern main asteroid belt.
The modern main asteroid belt SFD is therefore a poor model for producing the observed lunar highlands crater population.
This is due to the relative abundance within the main asteroid belt of objects that would produce lunar basins larger than Imbrium (and some larger than South Pole-Aitken) if they collided with the Moon, as compared with main belt objects that would produce \newtext{$D_c\simeq100\km$} craters.

From \newtext{Fig.}~\ref{f:combined_constraint} we show that the regional constraint is not met in any of our runs at \newtext{$\bar{N}_{pf}(D_i>10\km)<4\times10^{-6}\km^{-2}$.
Our best fit value to the regional constraint is $\bar{N}_{pf}(D_i>10\km)=5\times10^{-6}\km^{-2}$.}
However, at that level of cratering, very few of our runs satisfy the global constraint. 
\newtext{They nearly all produce at least one basin larger than Imbrium.
The average value of $\bar{N}_{pf}$ at the moment that a basin larger than $1200\km$ is produced is $\bar{N}_{pf}(D_i>10\km)=9.2\times10^{-7}\km^{-2}$.
}

A simple numerical exercise can be used to demonstrate why this is so by comparing the relative abundance of impactors that produce craters of these sizes within the main asteroid belt.
In \newtext{Fig.}~\ref{f:scalecompare} we plot the crater diameter vs. impactor diameter for craters produced in a {\em CTEM} global lunar surface simulation at \newtext{$\bar{N}_{pf}=5\times10^{-6}\km^{-2}$} for two different size ranges of crater.
We can see from this figure that to make a crater in the diameter range $90.5$--$128\km$ requires an impactor of $\sim3$--$15\km$, with most produced by impactors of $\sim5$--$6\km$ in diameter. 
To make a basin larger than the $\sim1200\km$ Imbrium basin requires an impactor of $\sim30$--$300\km$. 
On average an impactor with $D_i<70\km$ created at least one megabasin in these runs.

\begin{figure}[htb]
\centering
\includegraphics{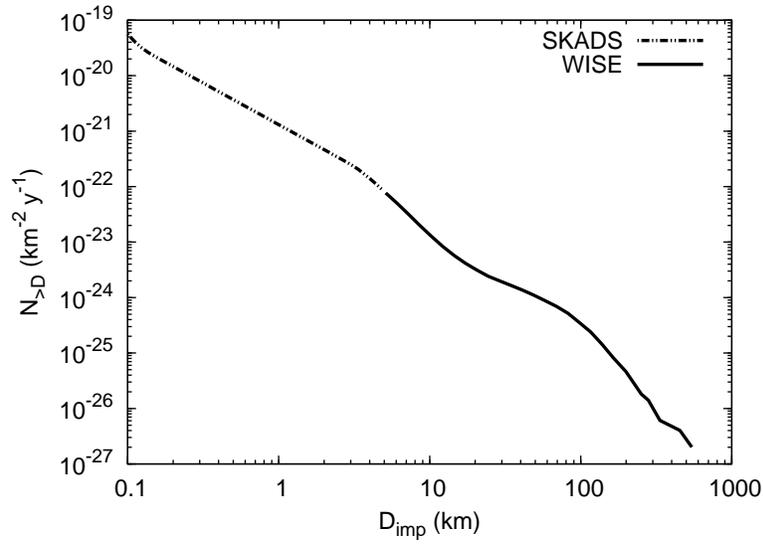}
\caption{Our impactor size frequency distribution is derived from the main asteroid belt. 
For $D_{i}>5\km$ we use the sizes of main belt asteroids from the Pass 1 WISE/NEOWISE data as reported by \cite{Masiero:2011jc} (solid line).
For the smaller asteroids we use the main belt SFD derived from a fit derived from the SKADS survey of \cite{Gladman:2009cx} (dash-dot line).
The SFDs are normalized such that the impact rate is similar to that used by the Neukum Production Function for the modern impactor flux \citep{Neukum:2001tl}.}
\label{f:MBA_CSFD}
\end{figure}

\begin{figure}[htb]
\centering
\includegraphics{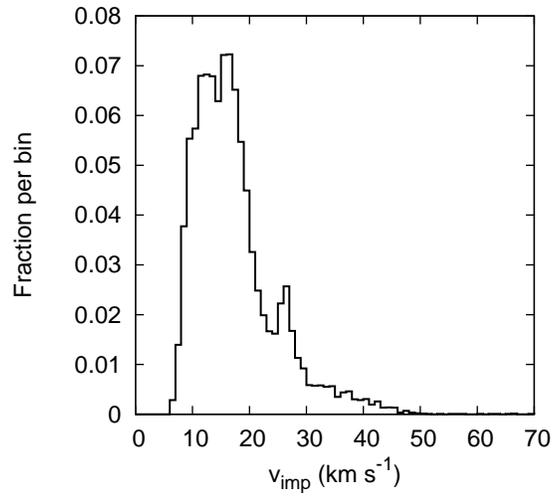}
\caption{Velocity distribution of impactors originating in the main asteroid belt onto the Moon based on N-body simulations~\citep{Minton:2010cn,Yue:2013bx}.}
\label{f:cehistory-impacts-Moon}
\end{figure}

\begin{figure}[htb] 
\centering
\includegraphics{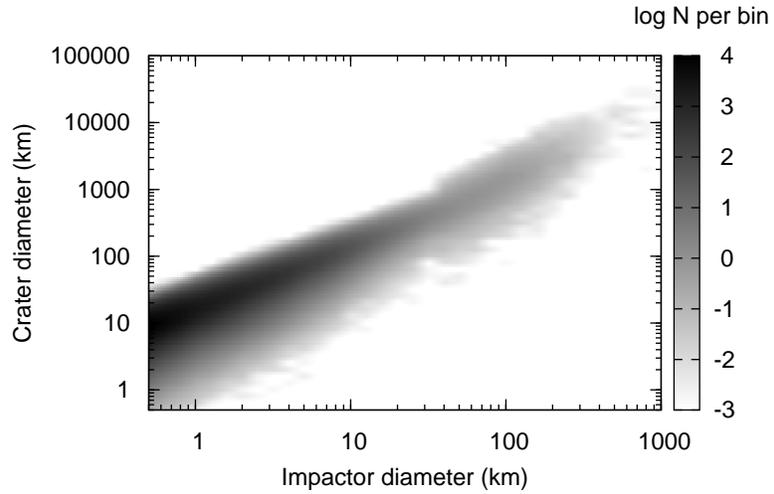} 
\caption{Example of the crater-size scaling relationship (from impactor to crater diameter) used in our models \citep{Holsapple:1993js}. 
The scaling relationship is for the Lunar surface with a surface gravity of \newtext{$1.62\m\s^{-2}$}, including the effects of our impact velocity distribution shown in Fig~\ref{f:cehistory-impacts-Moon}, and impact angle distribution given by Eq.~\ref{obliqueimpact_cumulative}.
We used $K_1=0.22$ and $\mu=0.55$ as our scaling parameters (see Eq.~\ref{volume_final}). 
The contours represent the average numbers of objects within bins of $0.1\times0.1$ log diameter (km) per simulation using $500$ simulations. 
}
\label{f:scalecontour}
\end{figure}

\begin{figure}[htb] 
\centering
\includegraphics[width=1.0\textwidth]{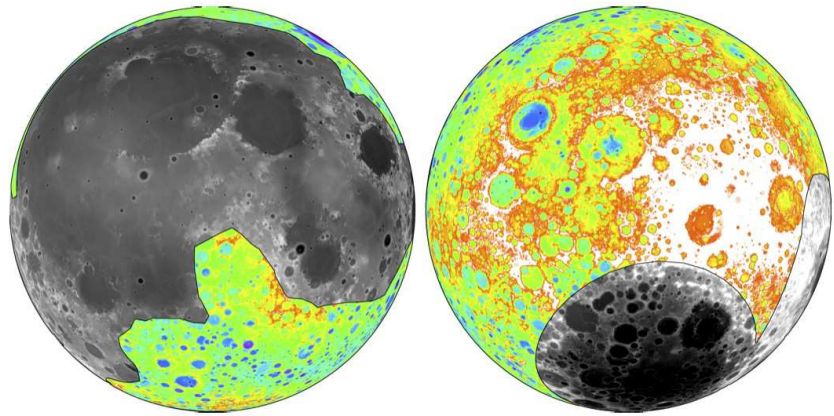} 
\caption{The color region is the terrain used in our regional simulations. This is the lunar highlands excluding SPA basin.}
\label{f:nonmarenoSPA}
\end{figure}

\begin{figure}[htb]
\centering
\includegraphics[width=0.49\textwidth]{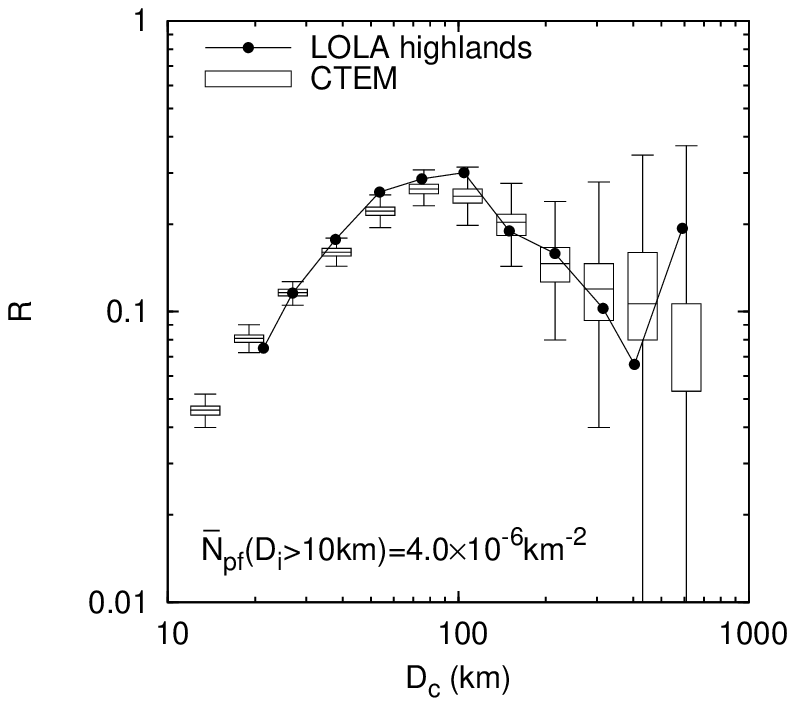}
\includegraphics[width=0.49\textwidth]{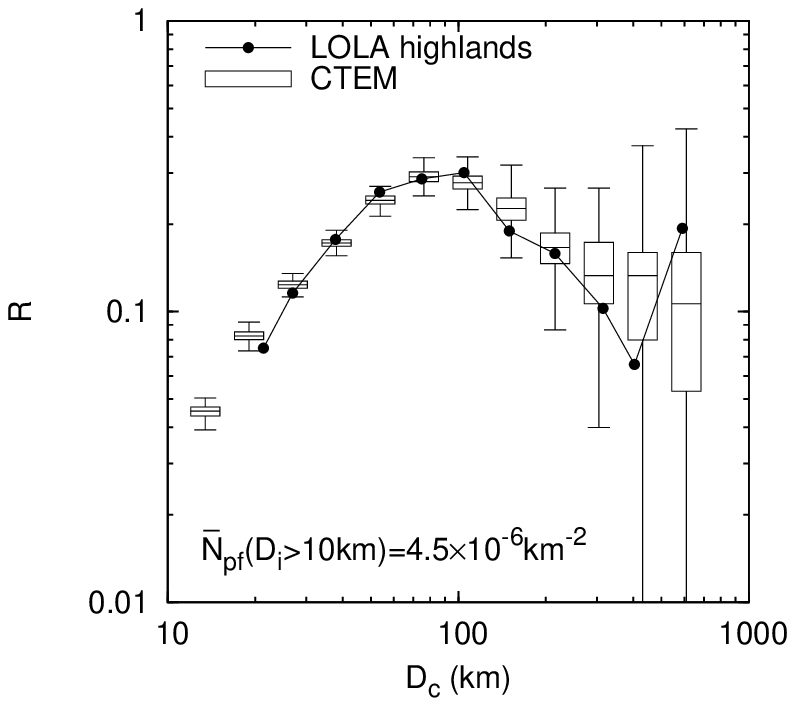}\\
\includegraphics[width=0.49\textwidth]{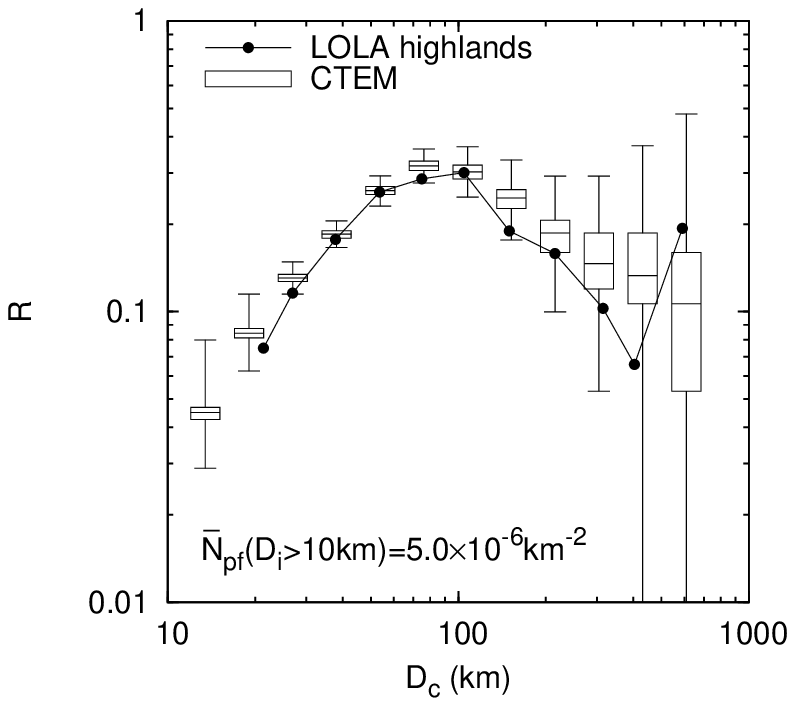}
\includegraphics[width=0.49\textwidth]{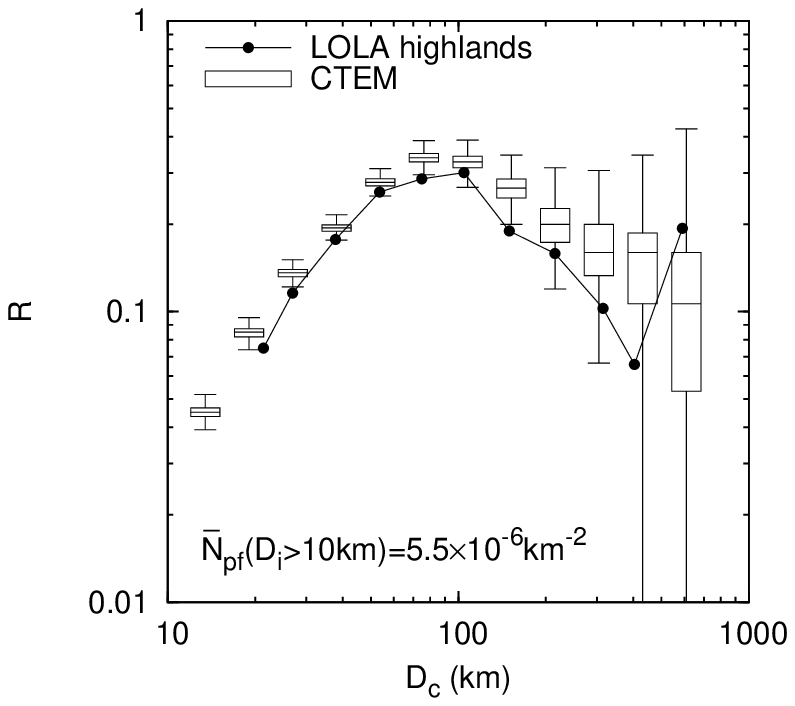}
\caption{R-plot comparison between the lunar highlands regional crater counts from the LOLA crater catalog and CTEM runs for four values of \newtext{$\bar{N}_{pf}(D_i>10\km)$: $4.0\times10^{-6}\km^{-2}$, $4.5\times10^{-6}\km^{-2}$, $5.0\times10^{-6}\km^{-2}$, and $5.5\times10^{-6}\km^{-2}$}.
The {\em CTEM} data is plotted as a box and whisker plot. 
The box is drawn to span the $25\%$ of data points above the median and the $25\%$ of data points below the median. 
The value of the median is drawn as a horizontal line within the box.
The error bars enclose $99\%$ of the data.
}
\label{f:regional-boxplot}
\end{figure}

\begin{figure}[htb]
\centering
\includegraphics{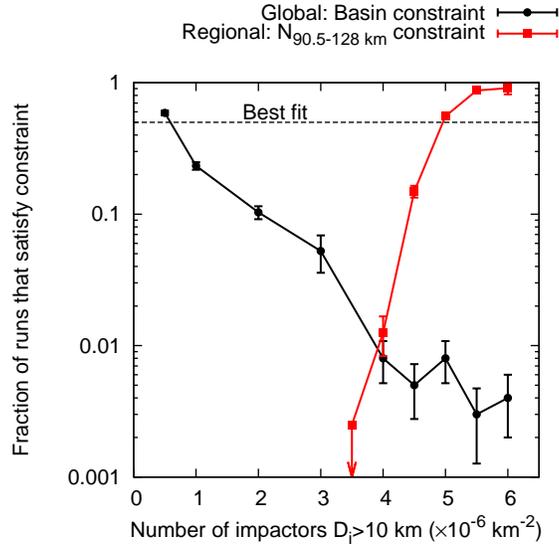}
\caption{Fraction of {\em CTEM} runs that two of our constraints as a function of \newtext{number of impactors, $\bar{N}_{pf}(D_i>10\km)$}. 
The circles are the global runs that satisfy the basin constraint: No more than $1$ basin with $D>1200\km$.
The squares are the regional runs that satisfy the $N_{90.5-128\km}$ constraint: The bin spanning $90.5$--$128\km$ must contain at least 198 craters, which is the number observed in the LOLA lunar highlands data set.
The impactors were drawn from the main asteroid belt size distribution. }
\label{f:combined_constraint}
\end{figure}

\begin{figure}[htb]
\centering
\includegraphics[width=0.49\textwidth]{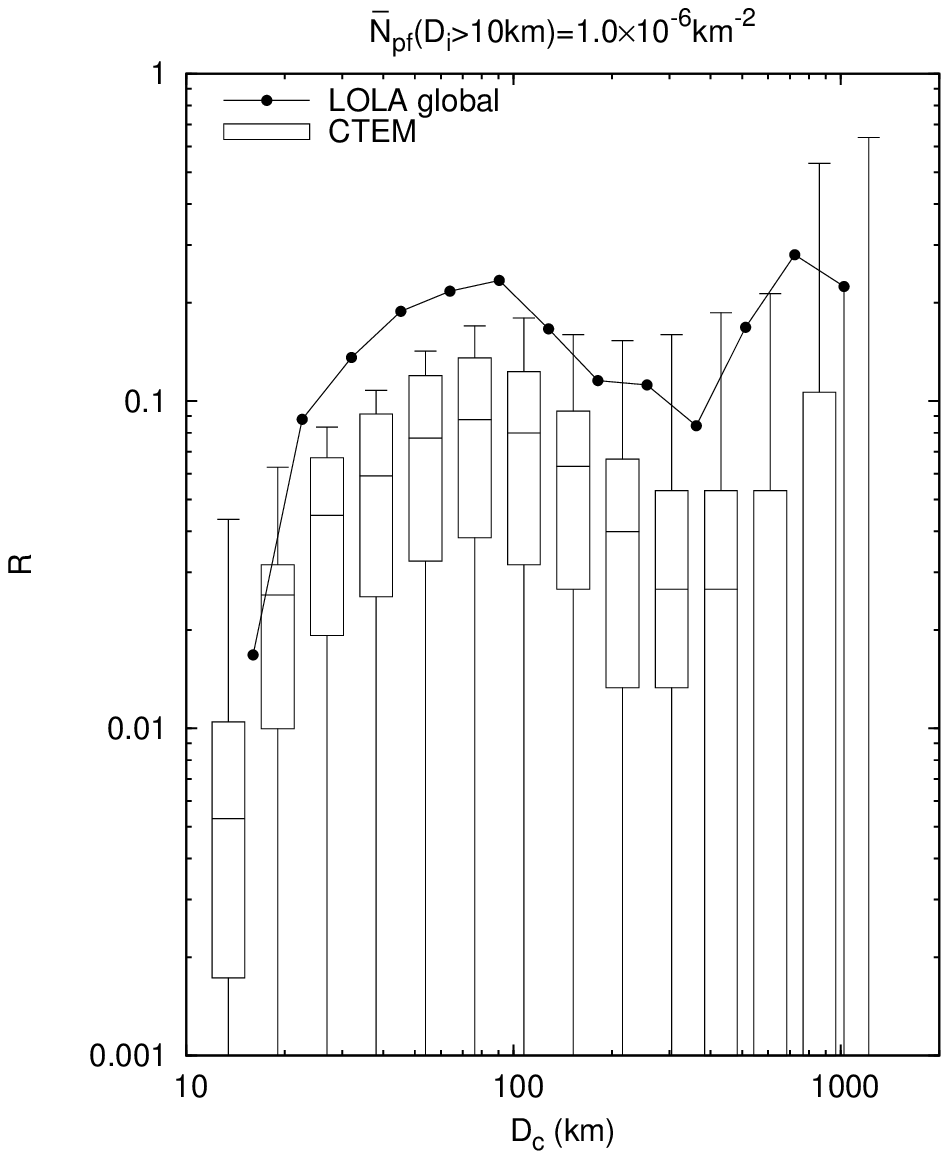}
\includegraphics[width=0.49\textwidth]{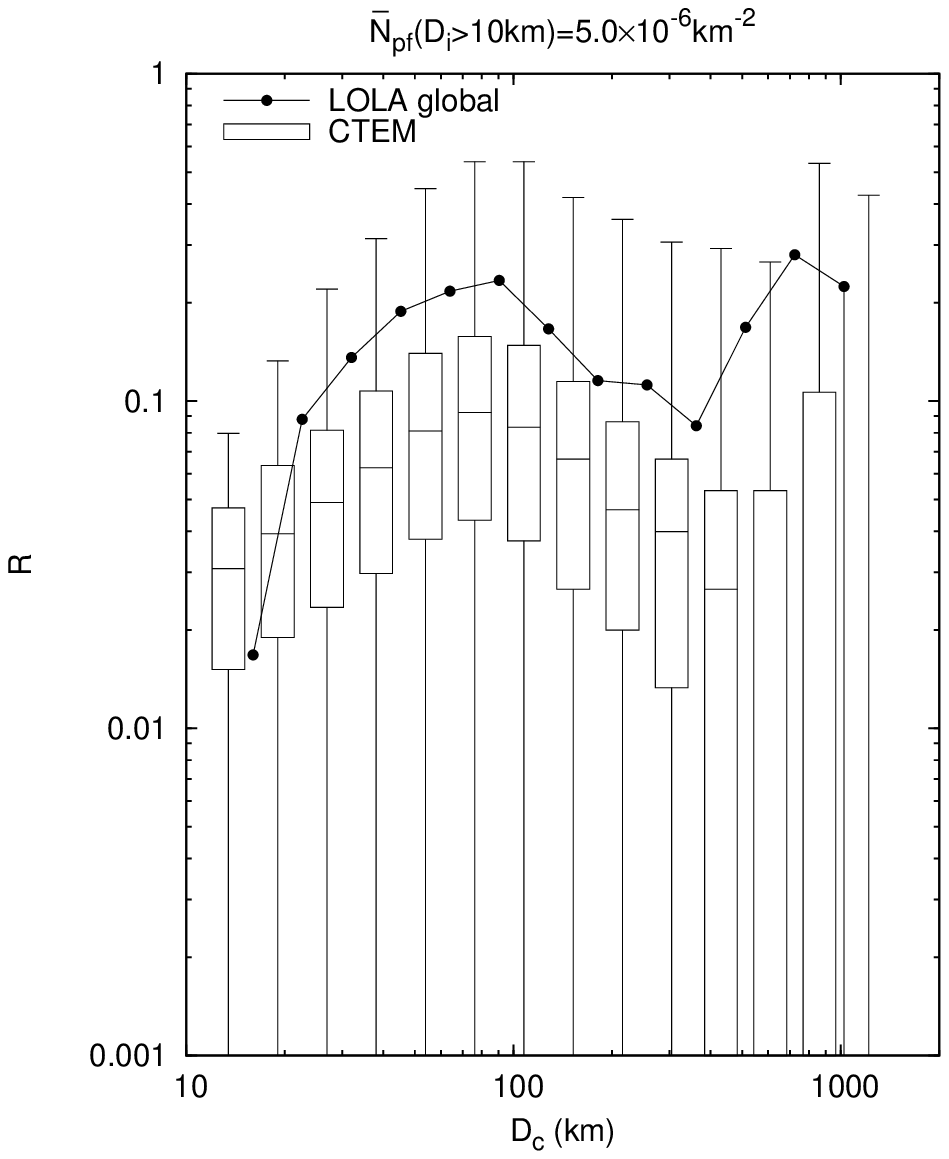}
\caption{R-plot comparison between the global crater counts from the LOLA crater catalog and {\em CTEM} runs with \newtext{$\bar{N}_{pf}(D_i>10\km)=1\times10^{-6}\km^{-2}$ (left) and $\bar{N}_{pf}(D_i>10\km)=5\times10^{-6}\km^{-2}$ (right)}. 
\newtext{In these runs, the simulations were ended when a basin larger than $1200\km$ diameter was generated, and therefore the average value of $\bar{N}_{pf}(D_i>10\km)=9.2\times10^{-7}\km^{-2}$.}
The {\em CTEM} data is plotted as a box and whisker plot. 
The box is drawn to span the $25\%$ of data points above the median and the $25\%$ of data points below the median. 
The value of the median is drawn as a horizontal line within the box.
The error bars enclose $99\%$ of the data.
\newtext{The left panel shows the global crater counts for the set of runs near the best fit value of $\bar{N}_{pf}$ for the global basin constraint.
The right panel shows the global crater counts for the set of runs near the  best fit value of $\bar{N}_{pf}$ for the regional $90.5<D_c<128\km$ constraint.}
}
\label{f:basincon-boxplot}
\end{figure}

\begin{figure}[htb]
\centering
\includegraphics{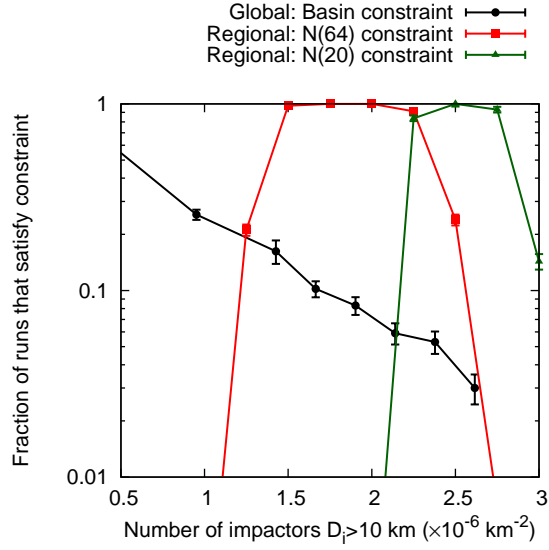}
\caption{ Fraction of {\em CTEM} runs that two of our E-belt constraints as a function of \newtext{$\bar{N}_{pf}(D_i>10\km)$}. 
\newtext{The black circles are the global runs that satisfy the basin constraints: No basins with $D_c>1200\km$.}
The red squares are the regional runs that satisfy the constraint that $N(64)=17\pm5$, and the green triangles are the runs that satisfy the constraint that $N(20)=135\pm14$, based on crater densities on Nectaris~\citep{Fassett:2012hr}.
The impactors were drawn from the main asteroid belt size distribution. 
}
\label{f:ebelt_constraint}
\end{figure}

\begin{figure}[htb]
\centering
\includegraphics[width=0.45\textwidth]{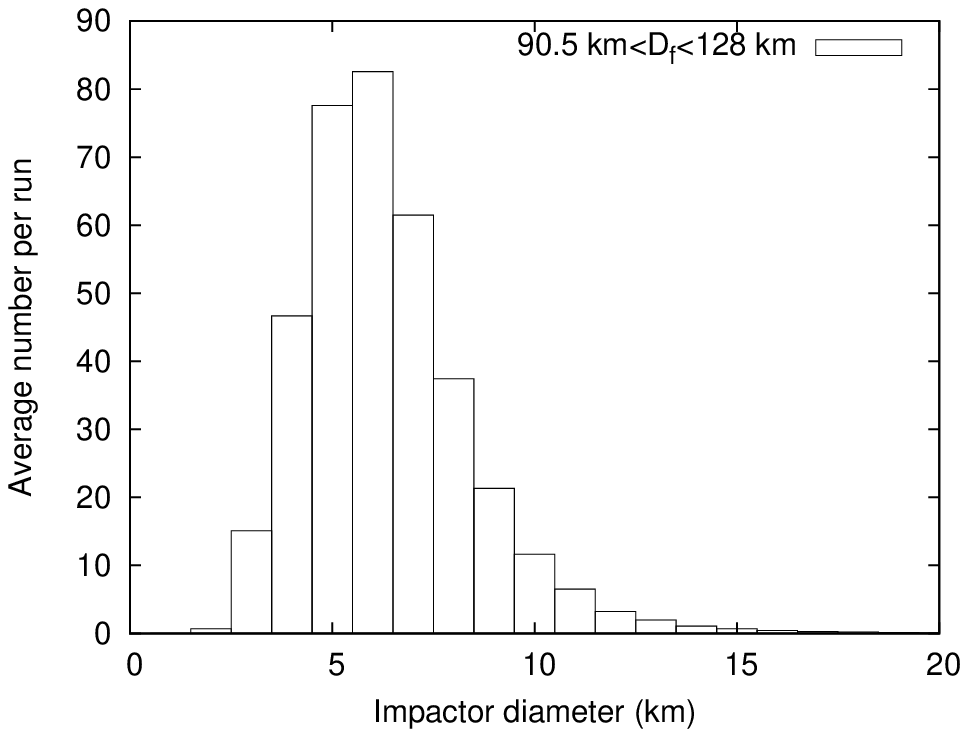}
\includegraphics[width=0.45\textwidth]{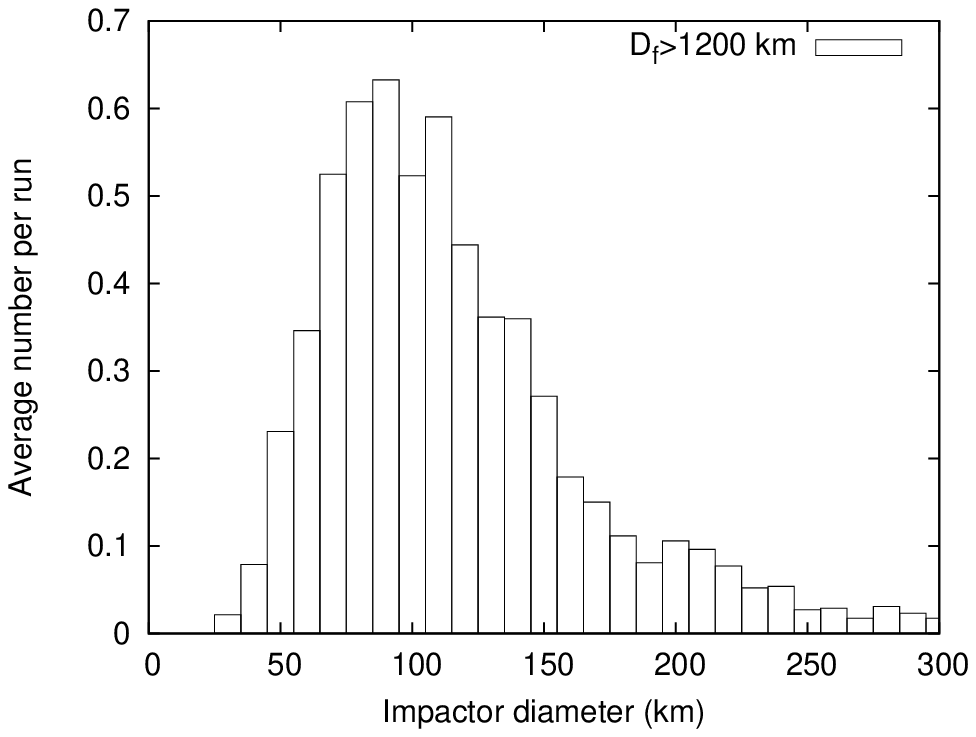}
\caption{Crater diameter vs. impactor diameter for craters produced in a {\em CTEM} global lunar surface simulation at \newtext{$\bar{N}_{pf}=5\times10^{-6}\km^{-2}$}.
The left panel shows the distribution of impactors of a given size (in bins of $1\km$) that produce craters with \newtext{$90.5<D_c<128\km$}.
The right panel shows the distribution of impactors of a given size (in bins of $10\km$) that produce craters with \newtext{$D_c>1200\km$}.
The plot shows the effect of all parameters that effect the scaling law in a given simulation, including impactor size, velocity, and impact angle, and in the case of impactors with $D_i>35\km$, our two crustal thermal profiles.
}
\label{f:scalecompare}
\end{figure}

In our LOLA crater catalog for the lunar highlands, we have $296$ craters larger than $90.5\km$ in diameter, and globally there is only one lunar basin larger than Imbrium basin, which is the SPA basin.
If we extrapolate the number of craters in the $2.0\times10^7\km^2$ area of highlands to reflect what the $3.8\times10^7\km^2$ global pre-mare lunar surface may have \newtext{experienced} we have $\sim560$ craters with \newtext{$D_c>90.5\km$}. 
Not every crater produced is observed due to the variety of crater erasure mechanisms described in Sec.~\ref{subsec:degradation}.
In our regional runs at \newtext{$\bar{N}_{pf}(D_i>10\km)=5\times10^{-6}\km^{-2}$}, we calculate that \newtext{$89\pm2.4\%$} of all craters produced with \newtext{$D_c>90.5$} are observable.
This gives us a ratio of \newtext{$N_{>90.5\km}/N_{>1200\km}\simeq630$}.
From the main asteroid belt model size-frequency distribution plotted in \newtext{Fig.}~\ref{f:MBA_CSFD}, we can estimate that the corresponding ratio of $N_{>5.5\km}/N_{>70\km}$ impactors is $\sim100$.
Therefore it is highly unlikely that the lunar surface could have been bombarded by enough main belt asteroids to produce the observed number of $D>90.5\km$ craters without producing many basins larger than Imbrium.
We quantified this probability in \newtext{Fig.}~\ref{f:combined_constraint}, which we find to be \newtext{$0.8\pm0.28\%$}.
We also showed in Section~\ref{subsec:ebelt} that the likelihood that the Nectarian crater densities could be produced while not producing any basins larger than Imbrium was \newtext{$\lesssim16\%$}.
We should note that we obtained this results assuming that the E-belt SFD was similar to the average main belt, rather than the inner main asteroid belt. 
However, the inner main asteroid belt may even more top-heavy than the main belt as a whole.
From \cite{Masiero:2011jc}, for the inner main belt $N_{>5.5\km}/N_{>70\km}=74$. 
\newtext{Using the inner main belt as the impactor population would therefore reduce the probability that the regional and global constraints could be met simultaneously.}

\begin{figure}[htb]
\centering
\includegraphics{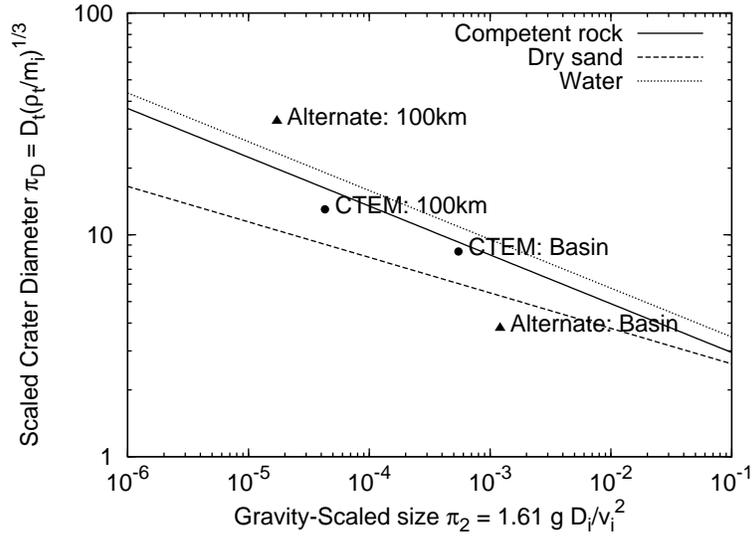}
\caption{ Gravity-scaled size $\pi_2=1.61 g D_i/v_i^2$ as a function of scaled dimensionless transient crater diameter $\pi_D=D_t\left(\rho_t/m_i\right)^{1/3}$.
The lines are the results of experimentally determined parameters for various materials~\citep{Wunnemann:2006jj}.
The circles show the scaling relationships used in this work for the typical impactors that generate the \newtext{$D_c=100\km$} craters ($D_i=5.5\km$) and \newtext{$D_c=1200\km$} megabasins ($D_i=70\km$); see also Fig.~\ref{f:scalecompare}.
The points are both calculated using the RMS value of our velocity distribution of $v_i=18.3\km\s^{-1}$. 
The triangles plot alternate scaling relationships that would be needed to easily satisfy both our global and regional constraints for the lunar highlands cratering record. 
The alternate for \newtext{$D_c=100\km$} (while keeping the basin scaling unchanged) requires $D_i=2.2\km$ impactor. 
The alternate for \newtext{$D_c=1200\km$} (while keeping the $100\km$ crater scaling unchanged) requires $D_i=156\km$.
Both of these alternative scaling relationships fall well outside of experimentally and numerically determined scaling relationships~\citep{Melosh:1989uq,Wunnemann:2003jg,Wunnemann:2006jj,Elbeshausen:2009dd}.
}
\label{f:impact_scaled}
\end{figure}

One possible solution to this problem could be that our scaling relationships between impactor size and final crater size are incorrect. 
If, for instance, the impactors that generated the \newtext{$D_c\simeq100\km$} craters were smaller (more numerous) than the $D_i\simeq5\km$ that we have estimated, or if the impactors that generated the \newtext{$D_c>1200\km$} megabasins were larger (less numerous) than the $D_i\gtrsim70\km$ that we have estimated, then it is possible that the main belt SFD could produce the required ratio \newtext{$N_{>90.5\km}/N_{>1200\km}=630$}.
Using our SFD, we estimate that in order to achieve this ratio, the impactors that generated the \newtext{$D_c\simeq100\km$} craters would have to be $D_i=2.2\km$, while keeping the basins scaling relationship unchanged.
Alternatively, the megabasin impactors would have to be $D_i=155\km$, while keeping the $100\km$ scaling relationship unchanged.
These values can be parameterized as a gravity-scaled impactor size, $\pi_2=1.61 g D_i/v_i^2$, and dimensionless transient crater diameter $\pi_D=D_t\left(\rho_t/m_i\right)^{1/3}$~\citep{Schmidt:1987vc,Melosh:1989uq,Wunnemann:2003jg,Wunnemann:2006jj,Elbeshausen:2009dd}.
We plot these alternate scaling results along with the experimentally and numerically derived scaling relationship of common materials in \newtext{Fig.}~\ref{f:impact_scaled}.
Both of these alternative scaling relationships fall well outside of experimentally and numerically determined scaling relationships.
Therefore it is unlikely that the scaling relationships for these craters are uncertain enough to explain our results.

One result of our study is the revelation that the small body population that cratered the lunar highlands was more depleted in $D>70\km$ objects than the main asteroid belt (or, equivalently, more \newtext{abundant} in $D>5\km$ objects).
The relative abundances of $D>70\km$ asteroids has been used as evidence that the initial size of planetesimals in the protoplanetary disk was this large~\citep{Morbidelli:2009dd}.
A similar feature in the size distribution is suspected for the Kuiper belt~\citep{Fraser:2009hk,Shankman:2013un}. 
In the case of the Kuiper belt, the ratio of these large objects to smaller objects is still quite uncertain, though work is ongoing to address this uncertainty~\citep{Richardson:2012uk,Minton:2012vm}.

There is certainly a great deal of similarity between the shape of the main belt size distribution and that of the population of impactors that produced the lunar highlands.
For instance, using our scaling relationship parameters of $\mu=0.55$ and $K_1=0.22$, which are within the range expected for impacts rocky objects onto rocky targets~\citep{Holsapple:1993js}, and our velocity distribution derived from main belt asteroids, the resulting craters both exhibit a peak on an R-plot near $100\km$ and a trough near $400$--$500\km$ (see \newtext{Fig.~\ref{f:regional-boxplot}}, for instance).
Caution must be used in over interpreting this similarity.
The critical specific energy required to disrupt a small body, $Q_D^*$, changes from strength-dominated for small objects to gravity dominated for large ones~\citep{Benz:1999cj}.
Bodies become weaker per unit mass as they become larger in the strength-dominated regime, but stronger as they become larger in the gravity-dominated regime.
This change in the size dependence of strength sets up standing waves in a collisionally evolved population, and the amplitude, wavelength, and phase of the waves depend on the shape of the strength law and the mutual collision velocities of the bodies in the population~\citep{OBrien:2003jk}. 
Therefore, the similarities seen between the lunar highlands impactors and the main asteroid belt \newtext{(when megabasins are ignored)} may simply reflect that these two populations had similar values of material strength (i.e. they were made of rock) and mutual impact velocities (i.e. they collided with each other while orbiting in the inner solar system).
\newtext{Nevertheless, we conclude that it is unlikely that the lunar highlands were bombarded by a population with a size-frequency distribution identical to that of the modern main asteroid belt.}

\section*{Acknowledgements}
The authors would like to thank H. Jay Melosh for his help in clarifying the crater scaling relationships. \newtext{Comments by Simone Marchi and an anonymous reviewer were most helpful in improving the paper, and we thank them for their thorough reviews. We would also like to acknowledge Bill Bottke and Renu Malhotra, who suggested additional tests that we incorporated into the paper.}
\clearpage
\bibliographystyle{elsarticle-harv}
%\bibliography{allrefs}

\clearpage

%%%%%%%%%%%%%%%%%%%%%%%%%%%%%%%%%%%%%%%%%%%%%%%%%%%%%%%%%%%%%%%%%%%%%%%%%%%%%
%% Tables 
%%%%%%%%%%%%%%%%%%%%%%%%%%%%%%%%%%%%%%%%%%%%%%%%%%%%%%%%%%%%%%%%%%%%%%%%%%%%%
\section*{Tables}

\newtext{
\begin{table}[htb]
\begin{center}
\caption{Calibration test surface properties}
\label{t:calibration-test}
\begin{tabular}{ l c c c c }
\hline
Simulation  & Grid size & Smallest counted crater & \# Craters & \# Craters\\
  Name      &    (px)   &  (cm)                   & Produced   & Counted \\
\hline
Terra Alta (6:1)    & $2000^2$ & $8.40$ & $496$  & $172$  \\
Terra Alta (6:1)    & $2000^2$ & $8.40$ & $969$  & $184$ \\
Mare Nostrum (16:1) & $2000^2$ & $4.15$ & $1468$ & $281$\\
Mare Nostrum (16:1) & $500^2$  & $2.05$ & $157$  & $74$ \\
Mare Nostrum (16:1) & $100^2$  & $1.01$ & $75$   & $28$ \\
\hline
\end{tabular}
\end{center}
\end{table}
}

\clearpage

%%%%%%%%%%%%%%%%%%%%%%%%%%%%%%%%%%%%%%%%%%%%%%%%%%%%%%%%%%%%%%%%%%%%%%%%%%%%%
%% Figures
%%%%%%%%%%%%%%%%%%%%%%%%%%%%%%%%%%%%%%%%%%%%%%%%%%%%%%%%%%%%%%%%%%%%%%%%%%%%%
\section*{Figures}

\thispagestyle{empty}

\clearpage

\thispagestyle{empty}

\clearpage

\thispagestyle{empty}

\clearpage

\thispagestyle{empty}

\clearpage

\thispagestyle{empty}

\clearpage

\thispagestyle{empty}

\clearpage

\thispagestyle{empty}

\clearpage

\thispagestyle{empty}

\clearpage

\thispagestyle{empty}

\clearpage

\thispagestyle{empty}

\clearpage

\thispagestyle{empty}

\clearpage

\thispagestyle{empty}

\clearpage

\thispagestyle{empty}

\clearpage

\thispagestyle{empty}

\clearpage

\thispagestyle{empty}

\clearpage

\thispagestyle{empty}

\clearpage

\thispagestyle{empty}

\clearpage

\thispagestyle{empty}

\end{document}